\documentclass[twocolumn,tighten,times]{aastex63}

\usepackage{amsmath}
\usepackage[caption=false]{subfig}

\newcommand{\Ms}{sMBH}
\newcommand{\Mp}{pMBH}
\newcommand{\Mtot}{$M_{\mathrm{bin}}$}

\newcommand{\fgd}{$f_{\mathrm{gd}}$}
\newcommand{\vg}{$v_{\rm g}$}
\newcommand{\vc}{$v_{\rm c}$}

\newcommand{\rhog}{$n_{\rm gd0}$}
\newcommand{\tevol}{$t_{\mathrm{evol}}$}

\newcommand{\D}{$D_{\rm ch}$}

\newcommand{\q}{$q$}

\newcommand{\dmax}{$d_{\rm ch}$}
\newcommand{\lmax}{$(L_2/L_1)_{\rm ch}$}

\newcommand{\fg}{$f_{\mathrm{g}}$}

\shorttitle{The Detectability of Kiloparsec Scale Dual AGNs}
\shortauthors{Li et al.}

\begin{document}

\title{The Detectability of Kiloparsec Scale Dual AGNs: The Impact of
  Galactic Structure and Black Hole Orbital Properties}

\author[0000-0002-0867-8946]{Kunyang Li}
\affiliation{School of Physics and Center for Relativistic
  Astrophysics, 837 State St NW, Georgia Institute of Technology,
  Atlanta, GA 30332, USA}   
\email{kli356@gatech.edu}

\author[0000-0001-8128-6976]{David R. Ballantyne}
\affiliation{School of Physics and Center for Relativistic
  Astrophysics, 837 State St NW, Georgia Institute of Technology,
  Atlanta, GA 30332, USA}
\email{david.ballantyne@physics.gatech.edu}

\author[0000-0002-7835-7814]{Tamara Bogdanovi{\'c}}
\affiliation{School of Physics and Center for Relativistic
  Astrophysics, 837 State St NW, Georgia Institute of Technology,
  Atlanta, GA 30332, USA}
\email{tamarab@gatech.edu}




\begin{abstract}
Observational searches for dual active galactic nuclei (dAGNs) at kiloparsec separations are crucial for understanding the role of galaxy mergers in the evolution of galaxies. In addition, kpc-scale dAGNs may
serve as the parent population of merging massive black hole (MBH)
binaries, an important source of gravitational waves. We use a
semi-analytical model to describe the orbital evolution of unequal
mass MBH pairs under the influence of stellar and gaseous dynamical
friction in post-merger galaxies. We quantify how the detectability of
approximately 40,000 kpc-scale dAGNs depends on the structure of their
host galaxies and the orbital properties of the MBH pair. Our models
indicate that kpc-scale dAGNs are most likely to be detected in
gas-rich post-merger galaxies with smaller stellar bulges and
relatively massive, rapidly rotating gas disks. The detectability is
also increased in systems with MBHs of comparable masses following low
eccentricity prograde orbits. In contrast, dAGNs with retrograde, low
eccentricity orbits are some of the least detectable systems among our
models. The dAGNs in models in which the accreting MBHs are allowed to
exhibit radiative feedback are characterized by a significantly lower
overall detectability. The suppression in detectability is most
pronounced in gas-rich merger remnant galaxies, where radiation
feedback is more likely to arise. If so, then large, relatively gas
poor galaxies may be the best candidates for detecting dAGNs.
\end{abstract}




\section{Introduction}
\label{sec:intro}
The hierarchal formation model of galaxy evolution predicts that
massive galaxies are built up through a series of mergers \citep[e.g.,][]{White1978,White1991}. As the nuclei of most massive galaxies contain a
massive black hole (MBH; e.g., \citealt{Kormendy2013}), it is
expected that many mergers will lead to a remnant with at least two
MBHs \citep[e.g.,][]{BBR1980}. Most mergers generate significant nuclear gas
flows \citep[e.g.,][]{D2005} that provide a favorable environment in which the
MBHs can accrete and shine as AGNs. Therefore, a population of dual
AGNs (dAGNs) in post-merger galaxies is an unavoidable prediction of
hierarchal galaxy formation \citep[e.g.,][]{Rosa2020}. In addition, dAGNs are expected to be the parent
population of binary MBHs, where the two MBHs are gravitationally
bound and the orbit decays through the emission of gravitational waves \citep[e.g.,][]{BBR1980,LISA2017,KBH2017,Kelley2019}. Thus, understanding the
population of dAGNs is necessary for determining the expectations for
future gravitational wave experiments.

Observational searches for dAGNs at radio, optical, infrared and X-ray
wavelengths have uncovered only a small number of
confirmed systems, especially those with separations $< 1$~kpc \citep[e.g.,][]{Foord2019,Gross2019,Hou2019,Foord2020,Hou2020,Rosa2020,Severgnini2021}. There are two principle difficulties in identifying kpc-scale dAGNs: (1) the small separation of dAGNs
(equivalent to $\sim 1$~arcsecond at
$z \approx 0.05$), requiring exquisite angular resolution, and (2) significant
obscuration \citep{Sco1986, Sar1987, Sar1989, Hop2005, Nara2008, Hop2012} in the nuclei of post-merger galaxies. As a result, despite significant
observational effort, the number of confirmed kpc-scale dAGN remains very low ($< 10\%$)  \citep{BS2011, Koss2012, Teng2012, Mez2014, Fu2015a, Fu2015b} with respect to the expectations based on the merger rate of galaxies \citep{Spring2005, Hop2005_m, VanW2012, Cap2017}, which limits the ability to compare to theoretical models and infer MBH binary merger rates.

Another set of challenges arises from models that attempt to predict
the formation and evolution of dAGNs \citep[e.g.,][]{Rosas2019}. Calculations of dAGN
formation in cosmological simulations often lack the resolution
needed to calculate the observational properties of the AGNs on
kpc or smaller scales. In contrast, simulations of isolated
merging galaxies can focus on the timescales and observational
properties of an evolving MBH pair in a merger remnant
\citep[e.g.,][]{Capelo2017}. These simulations are however relatively computationally expensive, and so only a limited number of merger configurations can be explored.

Theoretically, dynamical friction (DF) is expected to dominate the orbital decay of MBH
pairs from kpc separations until they are gravitationally bound. In
earlier work \citep[][hereafter LBB20a]{LBB2020}, we deployed a
semi-analytical model to study the effects
of galactic and orbital parameters on the inspiral time and
eccentricity evolution of a secondary MBH due to gaseous and stellar
DF at kpc scales. Here, we add a prescription to describe accretion onto both MBHs in our model and consider the observational properties of these dAGNs in $\approx 40,000$
simulations. In particular, we quantify the detectability of each dAGN
system as a function of three key parameters: inspiral time,
characteristic separation, and characteristic luminosity ratio. By
  considering how the detectability varies with galaxy
  (e.g., bulge mass, gas fraction) and orbital properties (e.g.,
  prograde vs.\ retrograde, high vs.\ low eccentricity), we are able to
  describe the types of post-merger galaxy
  remnants that are most likely to host detectable dAGNs.

This paper is organized as follows. Section~\ref{sec:methods}
summarizes the model for orbital evolution of a MBH in the remnant
galaxy due to DF and calculation of the AGN luminosities. In
Section~\ref{sub:prob}, we describe the probability distribution of
dAGN given in terms of their separations and luminosity
ratios. Section~\ref{sec:landeratio} presents the characteristic
luminosity ratios and separations of the model dAGNs and how they
depend on the properties of the host galaxy. The calculation of dAGN
detectability is also presented in this section. The implications of
our results for observational searches for dAGNs are discussed in
Section~\ref{sec:discuss}, with Section~\ref{sec:concl} providing the
conclusions.

\section{Methods}
\label{sec:methods}

Here, we provide a brief overview of the most important aspects
of our model and point the reader to LBB20a for a complete description. We then
introduce the method to calculate the accretion rate onto the two MBHs
and the emitted luminosity.

\subsection{Model of the Remnant Galaxy}
\label{sec:galaxymodel}

We assume that a galaxy merger produces a single remnant, with a stellar bulge and gas and stellar disk, which includes the MBH pair. The
primary MBH (\Mp; with mass $M_1$) is fixed at the center of the galaxy. The
non-rotating bulge has a mass $1000M_1$ \citep[e.g.,][]{M1998}, and
follows a power law density profile that is cutoff at the
characteristic outer radius \citep[e.g.,][]{BT1987}. We consider the orbital
evolution of a bare, secondary MBH (\Ms; with mass $M_2 < M_1$)
under the assumption that it has been stripped of any
gas and stars \citep[e.g.,][]{KBH2017}, and is orbiting in the plane of
the gas and stellar disks.  The total mass of the pair of MBHs is $M_{\mathrm{bin}}=M_1+M_2$ and the mass
ratio is $q=M_2/M_1$.

The gas disk in our model has an exponential profile with a scale radius $R_{\rm sd}$ \citep[e.g.,][]{BT1987}. The scale radius of the stellar disk is set to be $\log(M_{\rm 1}/10^{\rm 5} M_{\rm \odot})$ kpc, and that of the gas disk is two times $R_{\rm sd}$. Therefore galaxy models with larger \Mtot\ have gas densities that decrease more slowly with radius. The stellar bulge follows a coreless powerlaw density profile, with the scale parameters also proportional to $\log(M_{\rm 1}/10^{\rm 5} M_{\rm \odot})$ kpc. 

The gas and stellar disks rotate together with velocity $v_{\rm g}(r)$, defined in units
of local circular velocity $v_{\rm c}(r)$.  If the \Ms\ is co-rotating with the galaxy disks
on a prograde orbit, then we assign the disks' velocity as $v_{\mathrm{g}} > 0$, whereas in the case of a counter-rotating \Ms\  the disks' velocity is described as $v_{\mathrm{g}} < 0$. The
disks are further described by (a) the central gas number
density (\rhog), which determines the mass of
the gas disk within 1 kpc ($M_{\rm gd,1}$), and (b) the gas disk mass
fraction ($f_{\mathrm{gd}}=M_{\rm gd,1}/(M_{\rm gd,1}+M_{\rm sd,1})$), where $M_{\rm
  sd,1}$ is the mass of the stellar disk within 1\,kpc. Therefore, each merger remnant galaxy model is defined by five parameters (\Mtot, $q$, \rhog, \fgd\ and \vg), listed in
Table~\ref{tab:params}, which together yield a total of 39,366 model galaxies. 

The orbital evolution of the \Ms\ due to gaseous and stellar DF is followed until the separation between the two MBHs reaches $1$\,pc, at which point the simulation ends. In order to provide an intuitive description of orbit geometries, we characterize them in terms of the orbital eccentricity ($e$) and semi-major axis ($a$). As the orbits of the sMBH in the remnant galaxy are not closed, we use the farthest and closest approaches of the MBHs in each orbit to estimate $e$ and $a$. With these definitions, each simulation is initialized with $a \sim 1$ kpc and eccentricity $e_{\rm i}$. For presentation purposes, we focus on two groups of models: those with low initial eccentricity ($0 \le e_{\rm i} \le 0.2$) and high initial
eccentricity ($0.8 \le e_{\rm i} \le 0.9$). Thus, the suite of models spans a wide range
in initial orbital eccentricity and includes both prograde and
retrograde orbits.

\begin{deluxetable*}{ccC}
\tablenum{1}
\tablecaption{Galaxy Model Parameters\label{tab:params}}
\tablewidth{0pt}
\tablehead{
\colhead{Symbol} & \colhead{Definition} & \colhead{Values}
}
\startdata
 $v_{\rm g}(r)$ \ & gas disk rotational speed in units of $v_{\rm c}(r)$
 & -0.9$v_{\rm c}(r)$,\ldots,0.9$v_{\rm c}(r)$\ (\mathrm{step} = 0.1 v_{\rm c}(r)) \\
 $q$ & MBH mass ratio, $M_2/M_1$ & 1/n (n=2,\ldots,9) \\
 \Mtot\ & total MBH mass, $M_1+M_2$ & (2,3,5)\times 10^5 \mathrm{M}_{\odot}\\
 & & (1,3)\times 10^6 \mathrm{M}_{\odot}\\
  & & (1,3)\times 10^7 \mathrm{M}_{\odot}\\
  & & (1,3)\times 10^8 \mathrm{M}_{\odot}\\
 \rhog\ & gas number density at the center of galaxy & 100, 200, 300\, {\rm cm$^{-3}$}  \\
 \fgd\ & gas disk mass fraction, $M_{\rm gd,1}/(M_{\rm
   gd,1}+M_{\rm sd,1})$ & 0.3, 0.5, 0.9 \\
\enddata
\tablecomments{$v_{\rm g} < 0$ ($v_{\rm g} >0$) corresponds to the sMBH moving on a retrograde (prograde) orbit relative to the gas and stellar disk. $M_{\rm gd,1}$ and $M_{\rm sd,1}$ are the masses of the gas and stellar disks within a 1\,kpc radius, respectively. }
\end{deluxetable*}

\subsection{Dynamical Friction}
\label{sub:DF}

The orbital decay of the \Ms\ in our models occurs due to the combined effect of stellar
and gaseous DF. Stellar DF is exerted by both the bulge and the
stellar disk and is calculated using Eqs.~(5)-(7) in LBB20a, following
the work of \citet{AM2012}. The velocity distribution of stars in the
bulge is assumed to be Maxwellian (see equation\,(2) in \citealt{LBB20b}, hereafter
LBB20b). We assume that all stars in the stellar disk are rotating with a
speed $v_{\rm g}(r)$, so that the stellar velocity distribution is a
delta function defined at $v_{\rm g}(r)$. The contribution to the DF force from the stellar disk is negligible relative to the other galaxy components (LBB20a).

As gaseous DF depends on the Mach number of the moving body
\citep[e.g.,][]{O1999, KK2007}, the sound speed and thus, the
temperature of the gas disk must be defined for each model. We set the
temperature profile to be $10^4$\,K above the minimum temperature required by the Toomre stability criterion \citep{T1964}. This threshold effectively captures shock heating and turbulent
energy in the post-merger galaxy \citep[e.g.,][]{Barnes1991}. We calculate the gaseous DF force on the \Ms\ using Eqs.~(10)--(12) of LBB20a, which result in the gaseous DF force that is  strongest when the velocity difference between the \Ms\ and gas disk ($\Delta v$) is close to the sound speed, $c_s$ \citep{KK2007}.

According to LBB20a, the evolution time \tevol\ (time for the MBHs to reach a separation $< 1$~pc) can range from as short as $\sim 1$ Gyr if the stellar bulge dominates the evolution (or if the eccentricity grows large and the sMBH plunges below $1$ pc). However, if the gas disk dominates the DF force $t_{\rm evol}$ is more typically $\sim 5 $ Gyr, but can be as long as $\gtrsim 10$ Gyr depending on the orbital configuration of the sMBH.

\subsection{Calculation of Accretion Rate and Luminosity}
\label{sub:accretion}

The fixed \Mp\ and the moving \Ms\ are both allowed to accrete matter from their
surroundings and thus, may appear as a dAGN.  In the model outlined above, the \Ms\ may pass through
a wide range of environments during its orbital decay, speeding up and
slowing down as it sinks into the galaxy. Therefore, its luminosity as
an AGN is affected by both its motion and the properties of the host galaxy. 

The accretion rates onto both the \Mp\ and \Ms\ are calculated in each
simulation as a function of time. We describe the accretion rate onto the 
stationary \Mp\ using a constant Bondi accretion rate \citep{BH1944, Bondi1952} 
that depends only on the central properties of the model
galaxy. In addition, we assume the luminosity of the central AGN does
not exceed 10\% of the Eddington luminosity, in order to match the
typical values found in X-ray surveys of AGNs \citep[e.g.,][]{Lusso2012}. 
Specifically, the luminosity of the pMBH is determined by
\begin{eqnarray}
  \label{eq:l1}
L_{1} =
\begin{cases}
  0.1 \dot{M}_{\rm B1} c^2 & {\rm when} \;\;\; L_{1} < 0.1L_{\rm 1, Edd}, \\
  0.1 L_{\rm 1, Edd} & \mathrm{otherwise},
\end{cases}
\end{eqnarray}
where $\dot{M}_{\rm B1} = \pi\, n_{\rm gd0}\, m_{\rm p}\,(GM_{\rm 1})^2/c_{\rm s1,\infty}^3$ is the Bondi accretion rate of the primary and $L_{\rm 1, Edd} = 4\pi G M_{\rm 1}m_{\rm p}c / \sigma_{\rm T}$
%
%
%
%
is the Eddington luminosity of the pMBH. Here, $c_{\rm s1,\infty}$ is the sound speed of the unperturbed gas in the galaxy center, $\sigma_{\mathrm{T}}$ is the Thomson cross-section, and other constants have their usual meaning.  

Accretion onto the \Ms\ is calculated using the Bondi-Hoyle-Lyttleton \citep[BHL;][]{HL1939, BH1944, Bondi1952} model, where
accretion is no longer spherically symmetric since most of the inflow streams
past the MBH and is gravitationally focused on the symmetry axis of
the moving MBH. A fraction of the gas becomes gravitationally bound to
the sMBH and is accreted on it, giving rise to the luminosity
\begin{eqnarray}
  \label{eq:l2}
L_{2} =
\begin{cases}
  0.1 \dot{M}_{\rm BHL} c^2 & {\rm when} \;\;\; L_{2} < L_{\rm 2, Edd}, \\
  L_{\rm 2, Edd} & \mathrm{otherwise},
\end{cases}
\end{eqnarray}
where $L_{\rm 2, Edd} = 4\pi G M_{\rm 2}m_{\rm p}c / \sigma_{\rm T}$ is the Eddington luminosity for the sMBH. $\dot{M}_{\rm BHL} = \dot{M}_{\rm B2} / (1+\Delta v^2/c_{\rm s2, \infty}^2)^{3/2}$ 
%
%
is the BHL accretion rate of the secondary, with $\dot{M}_{\rm B2}$ representing the Bondi rate of the \Ms. $c_{\rm s2, \infty}$ refers to the sound speed of the unperturbed gas at the same galactocentric radius as the sMBH and $\Delta v$ is the velocity of the \Ms\ relative to the gas disk.  


The $\Delta v$ factor in $\dot{M}_{\mathrm{BHL}}$ leads to significant variations as the sMBH evolves through the galaxy (see Appendix A for a description of the time-averaged accretion rates). For example,
a smaller $\Delta v$ leads to a higher accretion rate onto the
\Ms, and the AGN luminosity will be largest when the velocity of the sMBH
is equal to the rotation speed of the galaxy, \vg. This implies that \Ms s on prograde orbits are more
luminous than those on retrograde orbit. However, AGNs on retrograde orbits 
with large eccentricity can also have large $L_2$ at the apocenter, where the \Ms\ has a fairly low $\Delta v$. Thus,  $L_2$ is affected by the distribution of gas in the galaxy remnant and can exhibit strong variations due to the orbital evolution of the \Ms.

The calculations described above assume that the MBH masses are constant in time. 
Similarly, the luminosities $L_1$ and $L_2$ are bolometric and do not account for any absorption or extinction within the merger remnant. To minimize the impact of these assumptions hereafter we characterize modeled dAGNs using the luminosity ratio, $L_2/L_1$, rather than individual AGN luminosities. This partly mitigates the impact of neglecting the intragalactic absorption and MBH growth. The impact of these assumptions is discussed in Section~5.2.



\section{Characterizing Dual AGN Properties}
\label{sub:prob}

\begin{figure*}[t]
  \centering
    \includegraphics[width=0.49\textwidth]{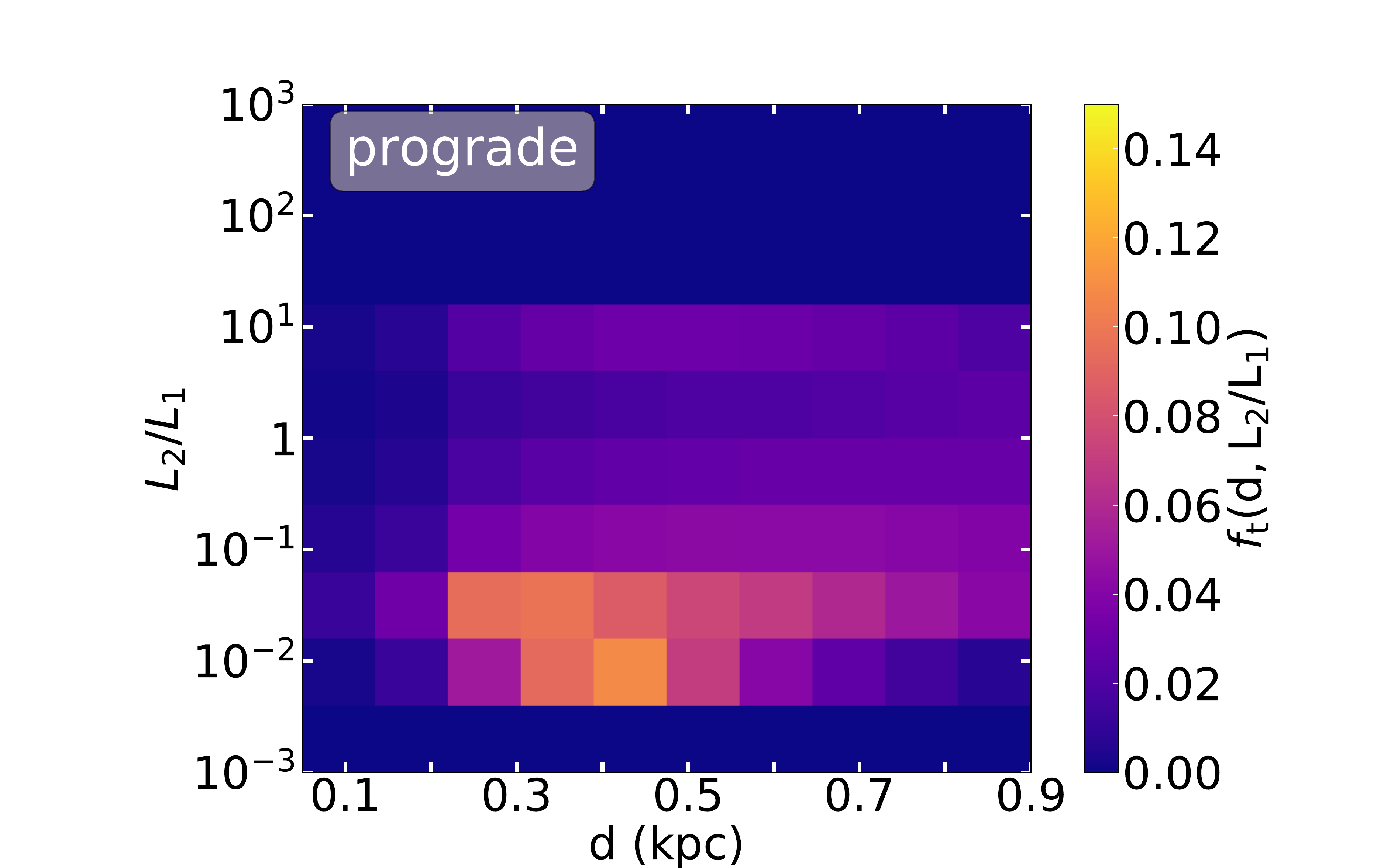}
    \includegraphics[width=0.49\textwidth]{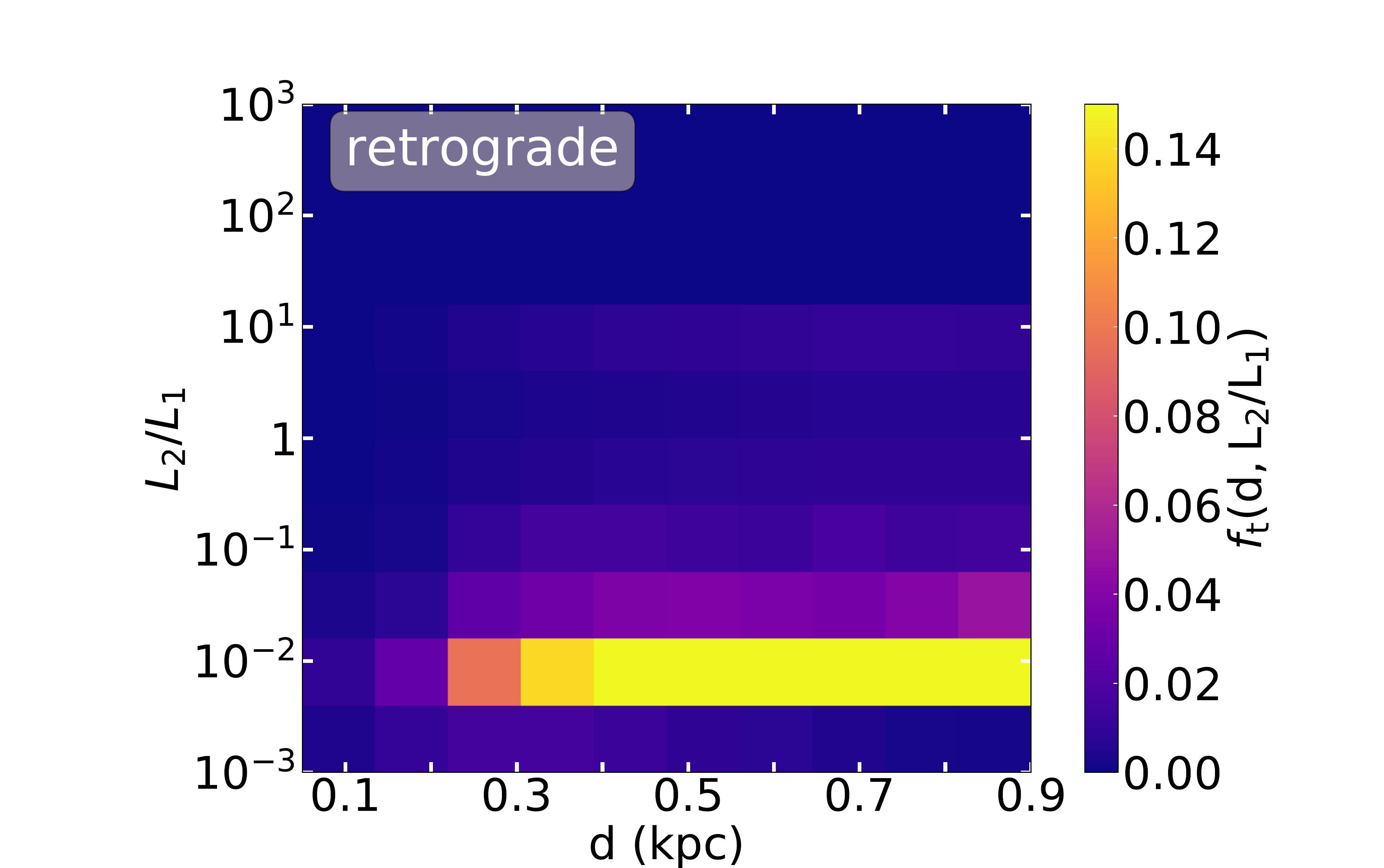}
\caption{The two-dimensional probability distribution of dAGNs on prograde (left panel) and retrograde orbits (right) in galaxies with $M_{\rm bin} = 3\times10^6 M_{\rm \odot}$, $q=1/9$,
and $v_{\rm g}=0.2 v_{\rm c}$. The distributions are summed over all values of \rhog\ and \fgd. The color bars mark the value of the probability.
}
\label{fig:dl_s}
\end{figure*}

\begin{figure*}[t]
\centering
  \includegraphics[width=0.49\textwidth]{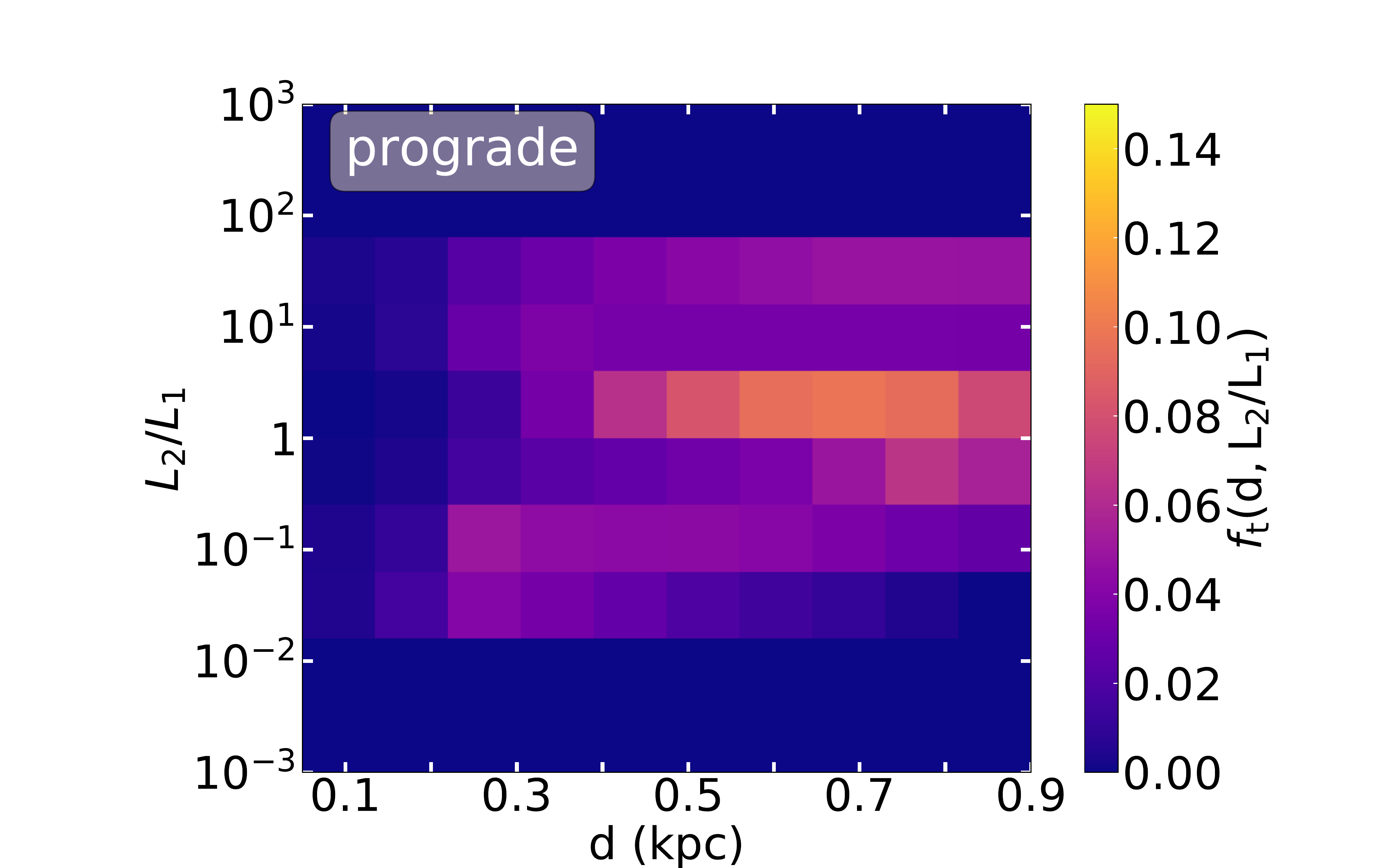}
  \includegraphics[width=0.49\textwidth]{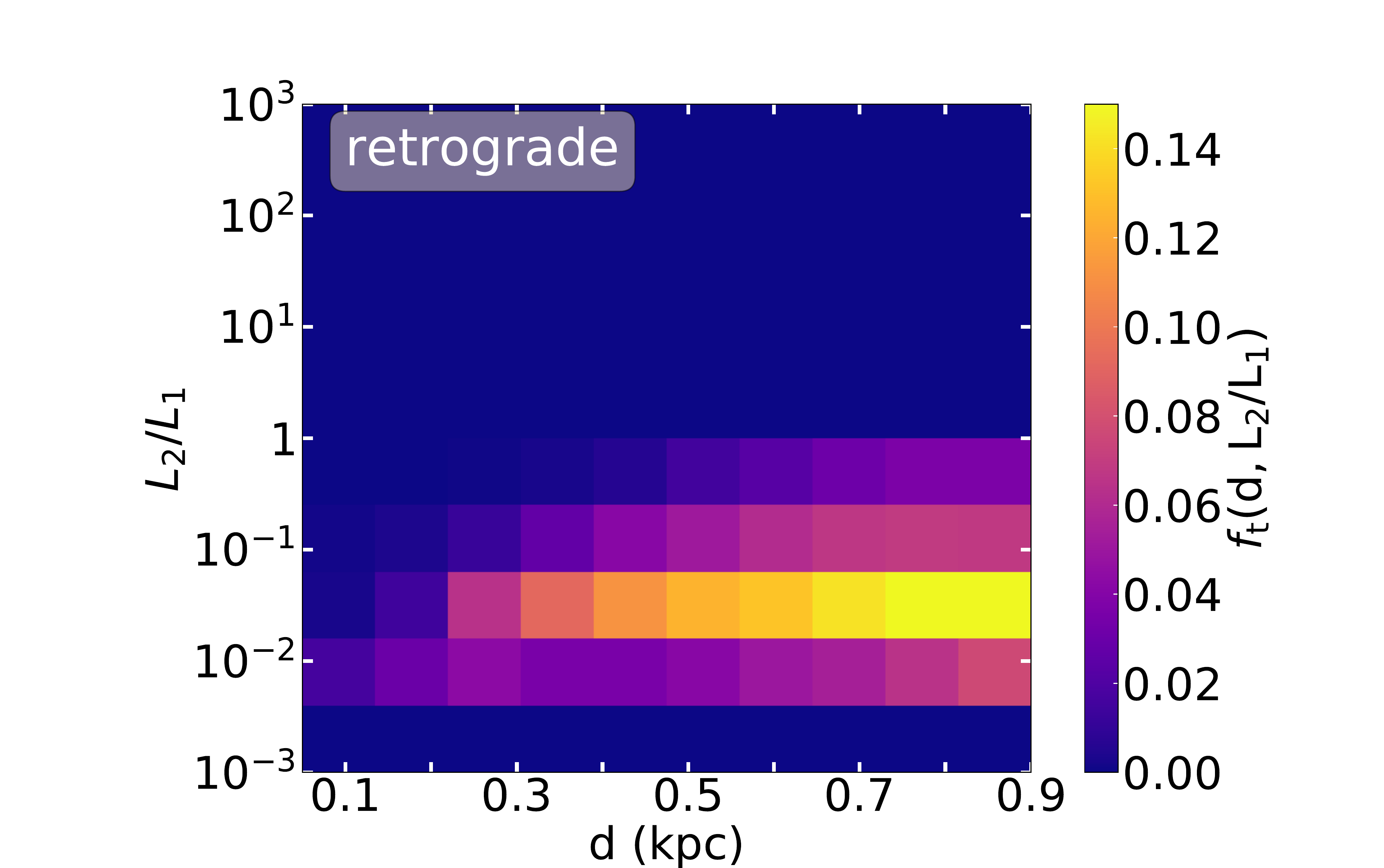}
\caption{Similar to Figure~\ref{fig:dl_s}, but for
  dAGNs with $M_{\rm bin} = 10^8 M_{\rm \odot}$, $q=1/3$, and
  $v_{\rm g}=0.8 v_{\rm c}$. 
 \ }
\label{fig:dl_l}
\end{figure*}

Observationally, the ability to successfully identify a kpc-scale dAGN depends on both its
separation ($d$) and luminosity ratio ($L_2/L_1$), where
larger values increase the chance of positive
detection\footnote{The separation $d$ used in this paper is
the physical distance between the two MBHs and not a projected distance on the sky. For a random distribution of orientations, the average projected separation is $(\pi/4)d$.}. It is therefore important to determine the conditions that could increase the chances of finding dAGNs.

Each of our models produces a description of the dAGN
position and luminosity over its evolution time, \tevol\ . 
In order to determine how the structure of the galaxy and
properties of the \Ms\ orbits affect the evolution of $d$ and $L_2/L_1$, we compute the
fraction of time that a dAGN spends at certain
$d$ and $L_2/L_1$.  For each of the 39,366 dAGN
simulations, we calculate a two-dimensional probability distribution, $f_t$, 
by summing the time the dAGN spends at a specific separation $d_i$ and luminosity ratio
$(L_2/L_1)_j$
%
%
\begin{equation}
  f_t\left [d_i, (L_2/L_1)_j\right ] = {1 \over
    t_{\mathrm{evol}}} \sum_{t} \Delta t \left[d_i, (L_2/L_1)_j \right],
\end{equation}
where $i=1,2,...,9$ and $j=1,2,...,6$. The range of separations ($0$, $0.9$ kpc) is evenly divided into nine bins with size of $0.1$\,kpc, while the logarithm of the luminosity ratio (ranging from $-3$ to $3$) is divided into six bins with a size 1. The maximum in the probability distribution occurs where the dAGN spends the largest fraction of time, and is identified as the `most-likely' combination of the luminosity ratio and separation at which the dAGN would be observed. 

\begin{figure*}[t]
\centering
\begin{tabular}{@{}cc@{}}
  \includegraphics[width=0.49\textwidth]{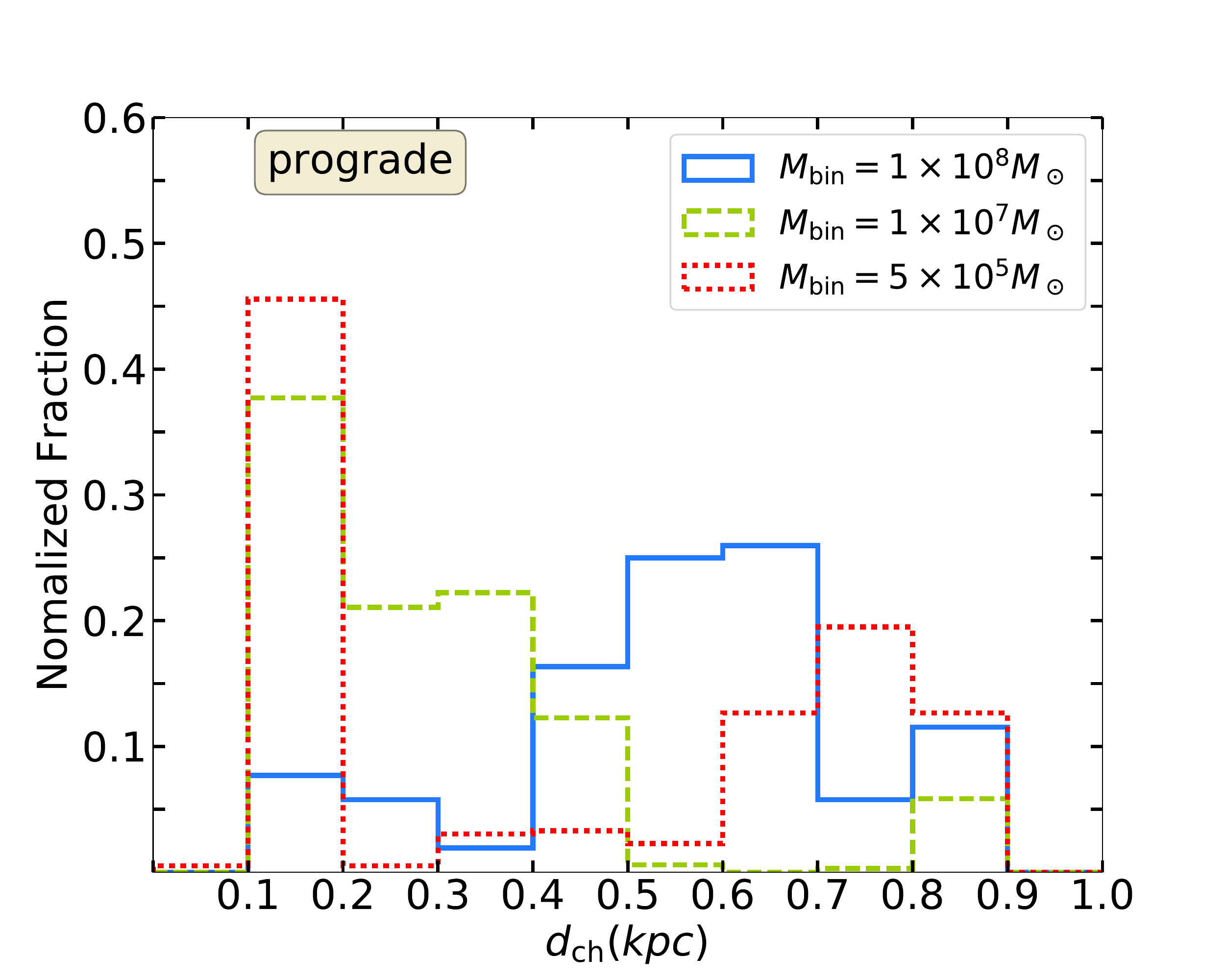}
  \includegraphics[width=0.49\textwidth]{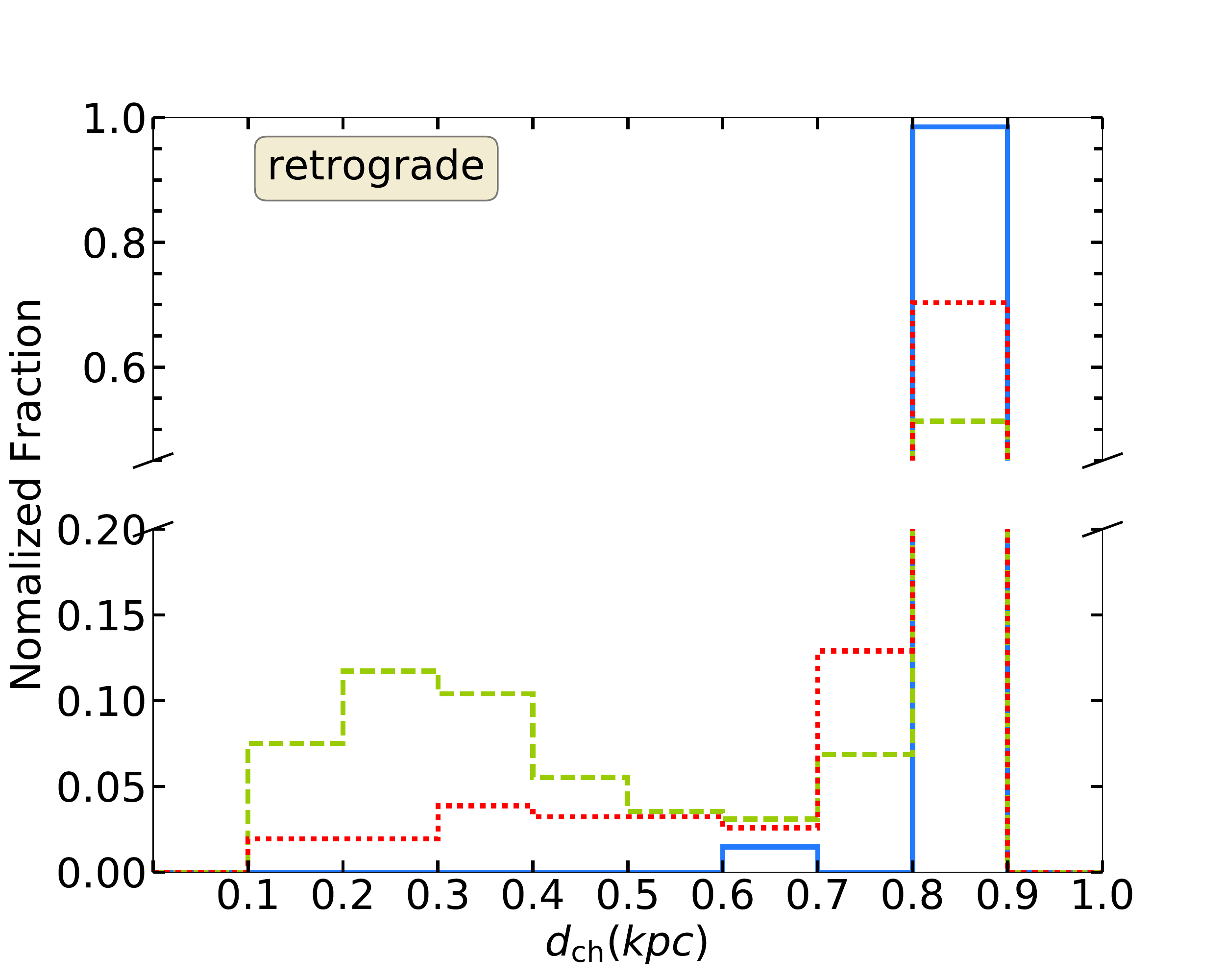}\\
\end{tabular}
\caption{Histograms of the characteristic (most probable) dAGN separation, \dmax\, for sMBHs on prograde (left) and retrograde (right) orbits and systems with different \Mtot.}
\label{fig:d_m}
\end{figure*}

Figures~\ref{fig:dl_s} and \ref{fig:dl_l} \footnote{See our Github page for the code and data for generating two-dimensional probability distribution of dAGNs: https://github.com/kli356/Dual-AGN-Detectability.} show examples of the probability distributions calculated in this way. In order to build up a signal that better highlights the impact of specific galactic or
orbital properties, both panels in the figures shows the sum of $f_t$
over all values of \rhog\ and \fgd\ while holding the remaining parameters ($M_{\rm
  bin}$, $q$ and $v_{\rm g}$) fixed.  Figure~\ref{fig:dl_s} shows $f_t$ for dAGNs with $M_{\rm
  bin} = 3\times10^6 M_{\rm \odot}$, $q=1/9$, and $v_{\rm g} =0.2
v_{\rm c}$ on prograde and retrograde orbits. The most likely luminosity ratio
for dAGNs with these properties is $\sim 10^{-2}$, and the most
probable dAGN separation is $d \sim 0.3-0.4$~kpc for prograde 
and $d \sim 0.4-0.9$~kpc for retrograde orbits. This difference
between the prograde and retrograde orbits is a natural consequence 
of the effect of gaseous DF. Due to the direction of gas dynamical friction at the apocenter of orbits, the eccentric prograde orbits decreases in eccentricity while moving through gas disks, while eccentric retrograde orbits become increasingly more eccentric. Therefore, eccentric prograde orbits tend to be circularized shortly after the start of the simulation, hence the separation between the two MBHs do not vary significantly during one orbit of evolution. However, the growth in eccentricity of eccentric retrograde orbits makes the variation of separation per orbit larger, which results in a wider range of potentially observed separations than that in eccentric prograde orbits (see LBB20a for details). 


Figure~\ref{fig:dl_l} shows $f_t$ for dAGNs in more massive and rapidly rotating galaxies, containing a comparable mass MBH pairs ($M_{\rm bin} = 10^8 M_{\rm \odot}$, $v_{\rm g}=0.8 v_{\rm c}$ and $q=1/3$).  Galaxies with these properties host dAGNs with larger
$L_2/L_1$ than those featured in Figure~\ref{fig:dl_s}. 
The higher luminosity ratio is due to the smaller relative
velocity between the \Ms\ and the rotating gas disk. Lower 
$\Delta v$ boosts the BHL accretion rate onto \Ms\ and
increases its luminosity. Similarly, in our model, more massive \Mp s 
reside in more massive and extended gas disks (Section~2.1), causing \Ms s in such systems to
experience larger gas densities (and consequently accretion luminosities) throughout their orbit. 
Larger $q$ values have similar effect on the accretion rate due to the more
massive \Ms s. Figure~\ref{fig:dl_l} also indicates that the most likely
dAGN separations are larger for MBH pairs in massive galaxies than for systems shown in the previous figure.  This is a direct consequence of the rate of evolution of these systems, which is relatively slow at large separations (where their evolution is gas disk dominated) and faster at smaller separation (where it is bulge dominated). Note that this trend is also present for lower mass galaxies shown in Figure~\ref{fig:dl_s} but that their smaller physical size implies smaller separations for dAGNs.  
Both Figures~\ref{fig:dl_s} and~\ref{fig:dl_l} indicate that the probability of detection of
dAGNs will depend on the type of galaxy in which the system
resides and the nature of the orbit of the \Ms.

\section{The Effect of Galactic and Orbital Properties on the Characteristics of Dual AGNs}
\label{sec:landeratio}

To determine how galaxy properties affect the most probable dAGN luminosity ratios and separations we sum the two-dimensional probability distributions described in the previous section along one of their axes. For example, the probability distribution for ($L_2/L_1$) is
\begin{equation}
  \label{eq:gt}
  g_t\left[ (L_2/L_1)_j \right] = \sum_{i} f_t\left [d_i, (L_2/L_1)_j \right ].
\end{equation}
The most probable luminosity ratio for dAGNs is given by the peak of this distribution. We thus define this maximum to be the characteristic luminosity ratio, \lmax, for a particular class of dAGNs with common properties,
\begin{equation}
  \label{eq:lchar}
  g_t \left[ (L_2/L_1)_{\mathrm{ch}} \right ] = \max \left [
    g_t \left [ (L_2/L_1)_j \right ] \right ].
\end{equation}
Similarly, the characteristic separation of a dAGN, \dmax, is defined as 
\begin{equation}
  \label{eq:dchar}
  g_t (d_{\mathrm{ch}}) = \max \left [ g_t(d_i) \right ],
\end{equation}
where $g_t(d_i)$ is the one dimensional probability distribution
calculated using an expression analogous to equation~(\ref{eq:gt}). We use these expressions to compute values of \lmax\ and \dmax\ for all dAGNs in our model suite.


\subsection{Dual AGN Separation}
\label{sec:d}

\begin{figure*}[t]
\centering
        \begin{tabular}{@{}cc@{}}
            \includegraphics[width=0.49\textwidth]{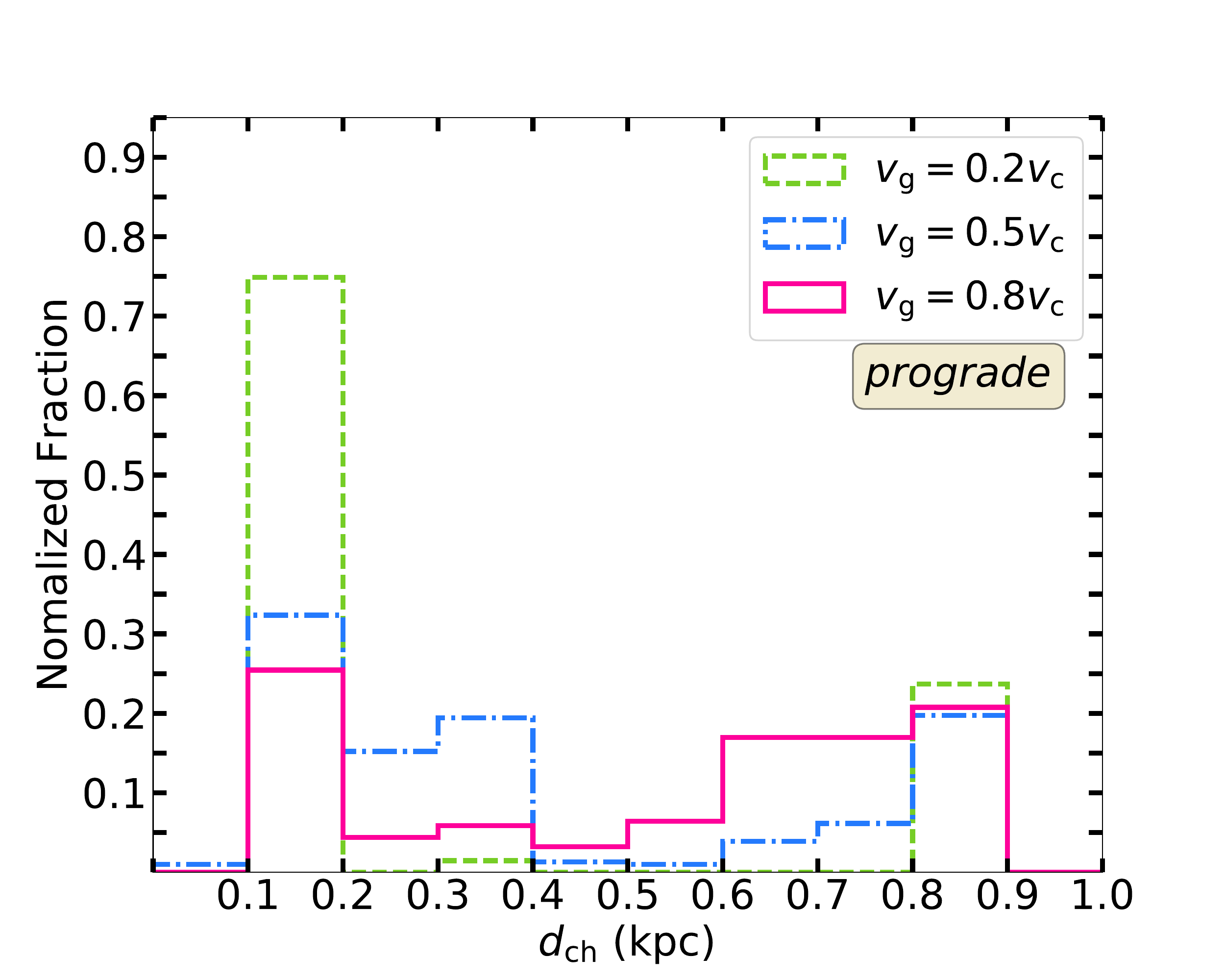}
            \includegraphics[width=0.49\textwidth]{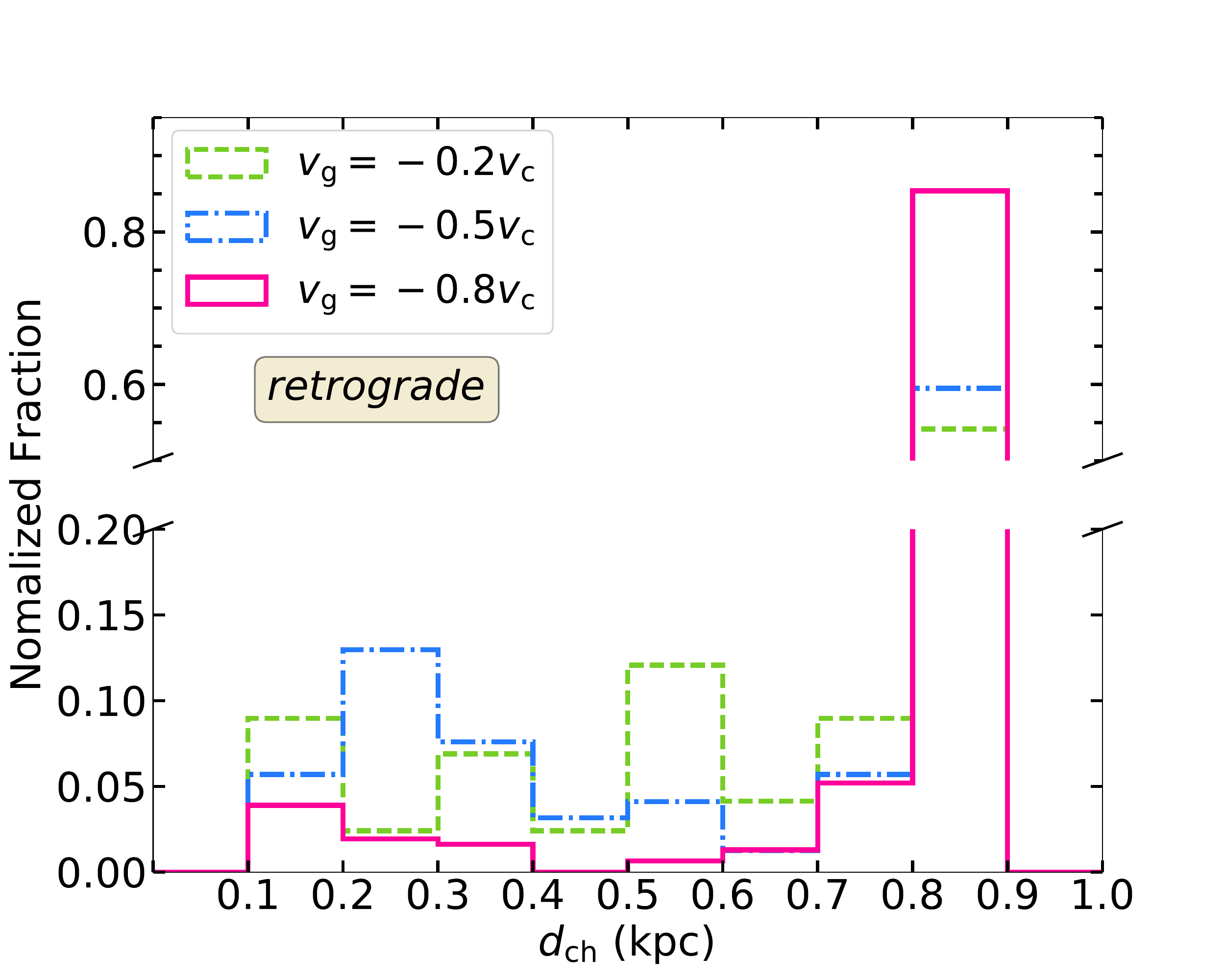}
        \end{tabular}
\caption{Histograms of the characteristic dAGN separation, \dmax\, for sMBHs on prograde (left) and retrograde (right) orbits as a function of the rotation speed of the merger remnant galaxy. Negative values of \vg\ indicate retrograde orbits.}
\label{fig:d_vg}
\end{figure*}

dAGNs are easiest to spatially resolve when they are widely separated on the sky. It is therefore important to understand which galaxy properties and orbital configurations increase the chances of observing a dAGN with a large separation. Figure~\ref{fig:d_m} shows histograms of the characteristic dAGN separations for different values of \Mtot, as well as the prograde or retrograde \Ms\ orbits. It illustrates that sMBHs on prograde orbits have smaller separations than those on retrograde orbits, on average. This is because the \Ms\ on prograde orbits tend to circularize, even if they started on eccentric orbits, whereas those on retrograde orbits tend to grow more eccentric (LBB20a). As a result, retrograde sMBHs are more likely to be observed with large, $\sim $~kpc separations.

Another important parameter that affects the distribution of \dmax\ is \Mtot. Regardless of the sense of rotation of the \Ms, the pairs with larger \Mtot\ have larger characteristic separations, particularly for prograde orbits. This can be understood because the mass and size of the bulge, as well as the gas and stellar disk of the remnant galaxy, are proportional to $M_1$ (see Section~\ref{sec:galaxymodel}). Thus, increasing
\Mtot\ for a fixed value of $M_2$ implies higher gas and stellar
density at a given radius in the remnant galaxy, and consequently,
faster orbital evolution of the \Ms. This means that the
characteristic separation where dAGNs spend a significant fraction of their time, and are likely to be observed, moves outward (toward larger values of \dmax) with increasing \Mtot.

We also examine the distribution of \dmax\ as a function of the rotation speed of the galactic disk and show results in Figure~\ref{fig:d_vg}. This is relevant because the magnitude of the DF force directly depends on the relative velocity between the \Ms\ and the rotating gas disk through which the MBH is moving (see Section~\ref{sub:DF}). For MBHs on prograde, and in general eccentric orbits, the cumulative effect of the gas DF force is largest at the apocenter, where the \Ms\ spends the largest fraction of its orbital period. For slowly rotating gas disks (i.e., $v_g = 0.2v_c$), the relative velocity at the apocenter for sMBHs on moderately eccentric orbits leads to increased efficiency of gaseous DF. Consequently, these orbits evolve quickly on average, and their dAGNs are most likely be observed with smaller separations, as illustrated in Figure~\ref{fig:d_vg}. As $v_g$ increases, the gaseous DF force acting on prograde sMBHs with moderate eccentricity at their apocenter diminishes. As a result, dAGNs in rapidly rotating galaxies  evolve more slowly and are more likely to be observed with \dmax$\sim 0.7$~kpc. When it comes to \Ms s on retrograde orbits, the relative velocity between the secondaries and their galactic gas disks is always large, and hence, these dAGNs are likely to be observed with large separations ($d_{\mathrm{ch}} \sim 0.8$--$0.9$~kpc), as shown in the right panel of Figure~\ref{fig:d_vg}.

In summary, for the \Ms\ on a prograde orbit, the most likely dAGN separation can be as low as $\sim 0.1$~kpc, especially if the merger galaxy is rotating slowly or has a low mass. We expect dAGN separations with the \Ms\ on a retrograde orbit to be $d_{\mathrm{ch}} \sim 0.8$~kpc, for a wide range of remnant galaxy properties. We also find that other properties of the merger remnant galaxies (e.g., \rhog) or MBH pairs (e.g., $q$) do not significantly affect the distribution of \dmax.

\subsection{Luminosity Ratios of Dual AGNs}
\label{sub:lratio}

\begin{figure*}[t]
\centering
        \begin{tabular}{@{}cc@{}}
            \includegraphics[width=0.49\textwidth]{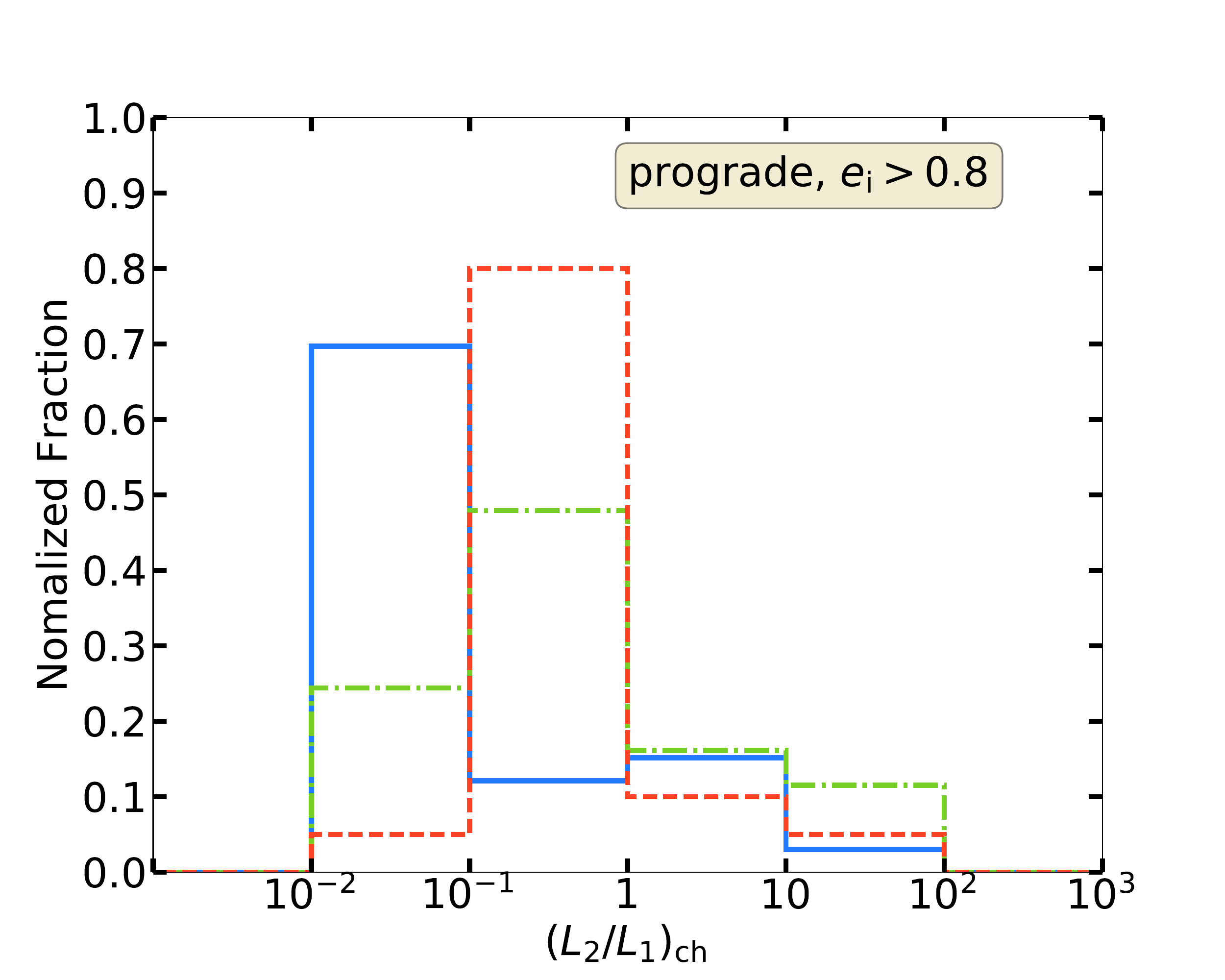}
            \includegraphics[width=0.49\textwidth]{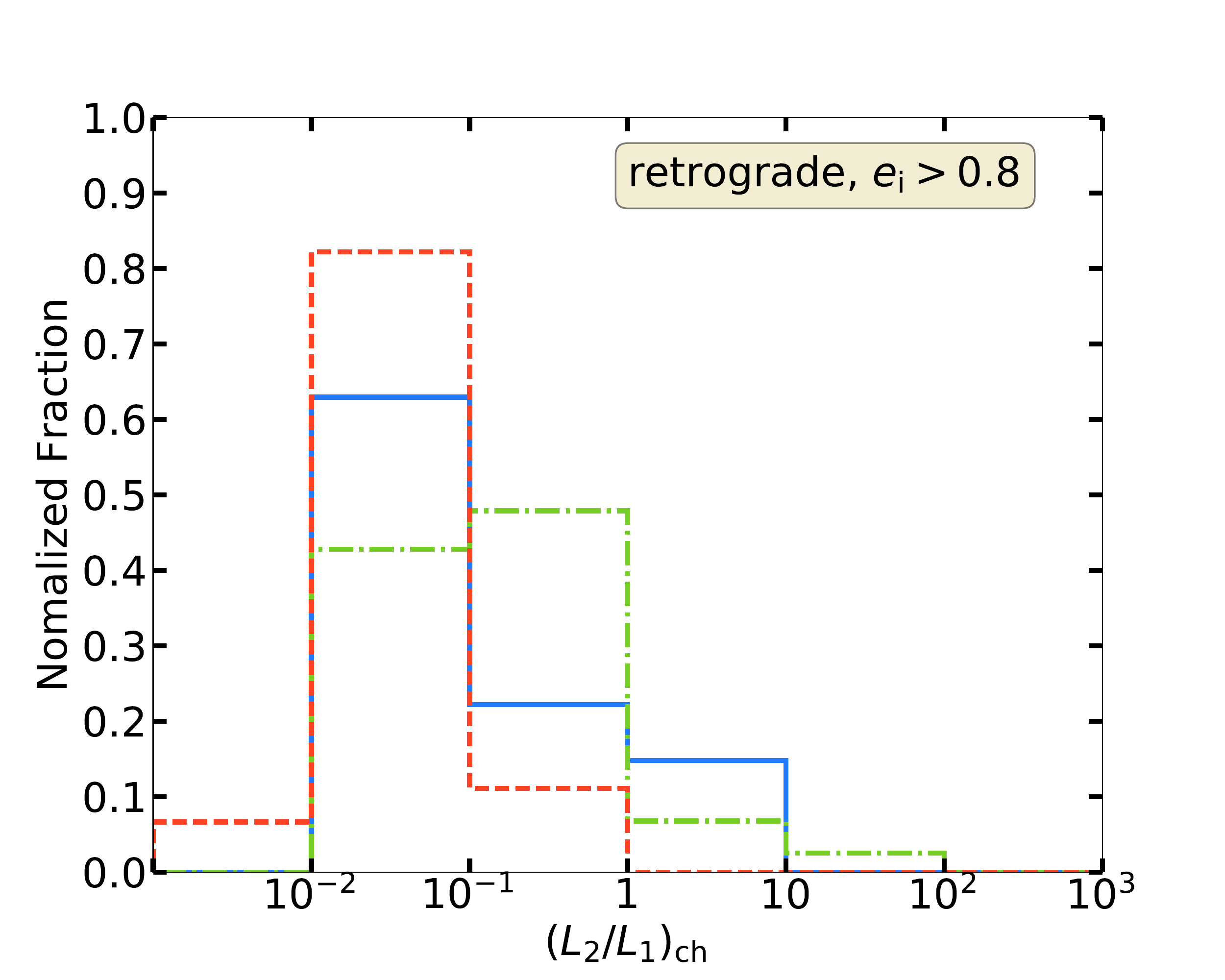}\\
            \includegraphics[width=0.49\textwidth]{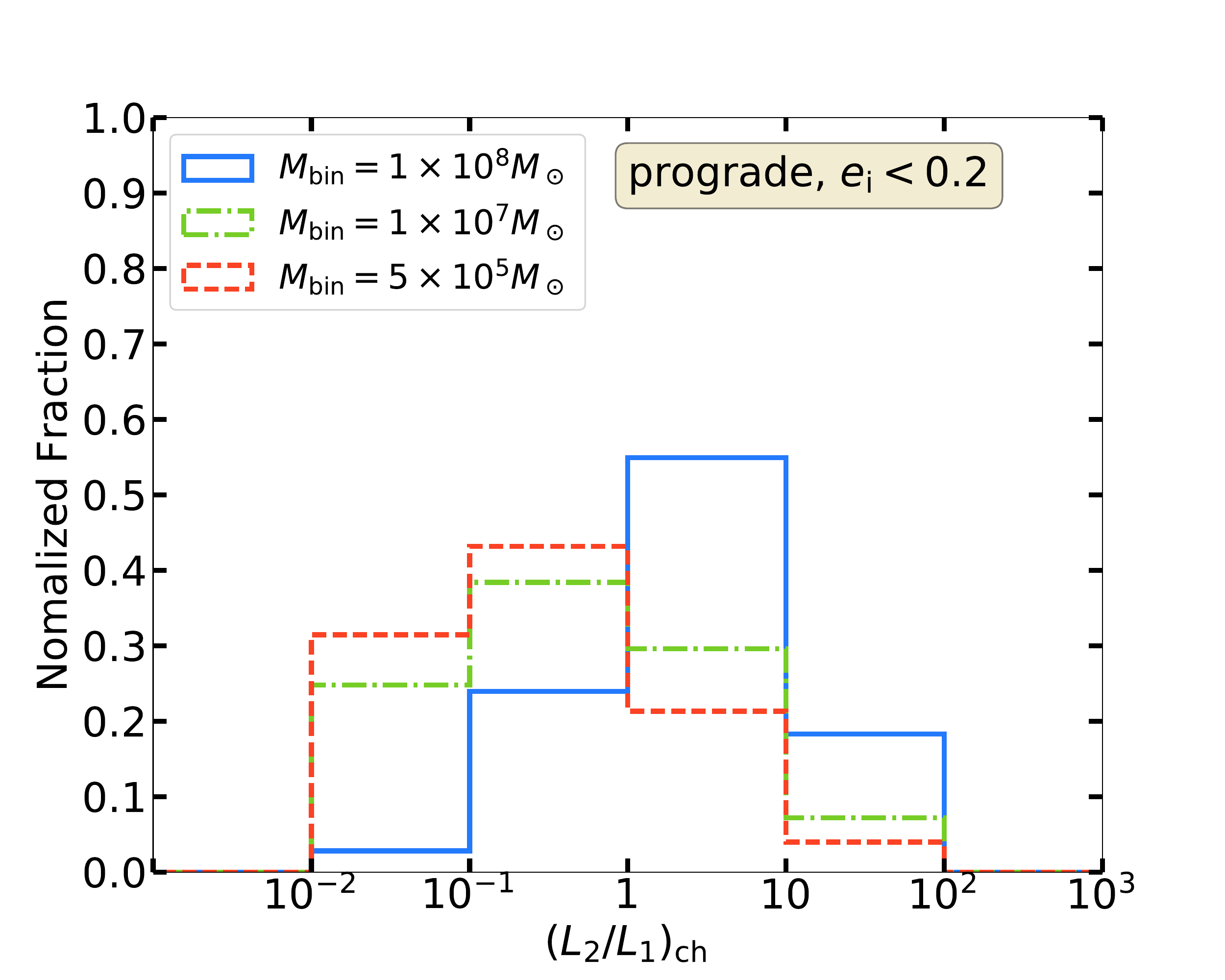}
            \includegraphics[width=0.49\textwidth]{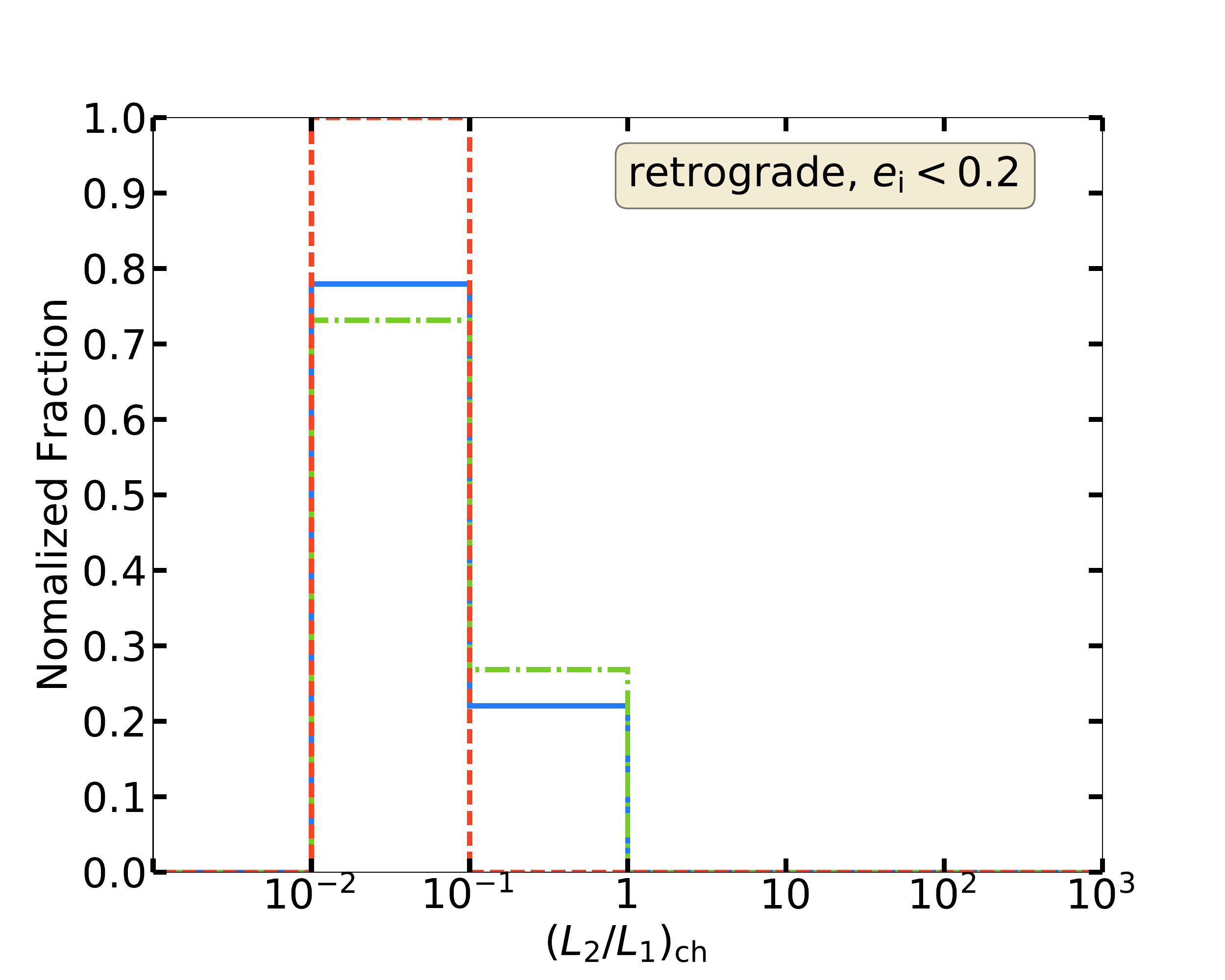}
        \end{tabular}
\caption{Histograms of the characteristic dAGN luminosity ratios, \lmax\, for \Ms\ on prograde (left) and retrograde (right) orbits of high (top) or low (bottom) initial eccentricity. Each panels shows the distribution of \lmax\ for different \Mtot.}
\label{fig:l_m}
\end{figure*}

In this section we discuss the distribution of the most likely, or characteristic, dAGN luminosity ratio \lmax. Figure~\ref{fig:l_m} shows the distribution of \lmax\ for different types of \Ms\ orbits (including the sense of rotation and initial eccentricity) as well as for different values of \Mtot. Since in our models the luminosity of the \Mp\ is constant in time in each individual merger scenario, the differences in \lmax\ are driven by the changes in accretion luminosity of the \Ms.  

For example, each panel illustrates that a larger \Mtot\ is more likely to produce higher values of \lmax. 
This is because the BHL accretion rate onto the \Ms\  is proportional to the local gas density, and the
gas disk in our model has an exponential profile with a scale radius that depends on ${\rm log}(M_{\rm 1})$. Hence, models with a larger \Mp\ have a gas density profile that decreases more slowly with radius
and the resulting \Ms\ accretion rates are larger. The effects of the radial gas density profile are also seen in the panels showing prograde orbits with different $e_i$. Very eccentric orbits bring the \Ms\ to the outskirts of the gas disk where the gas density is low. As a result, these dAGNs have significantly smaller \lmax\ compared to low $e_i$ systems. A \Ms\ on a retrograde orbit will have a large velocity relative to the gas disk ($\Delta v$) regardless of $e_i$, which will reduce its BHL accretion rate. As a result, \lmax\ values of these dAGNs are often less than 0.1.

\begin{figure*}[t]
            \includegraphics[width=0.49\textwidth]{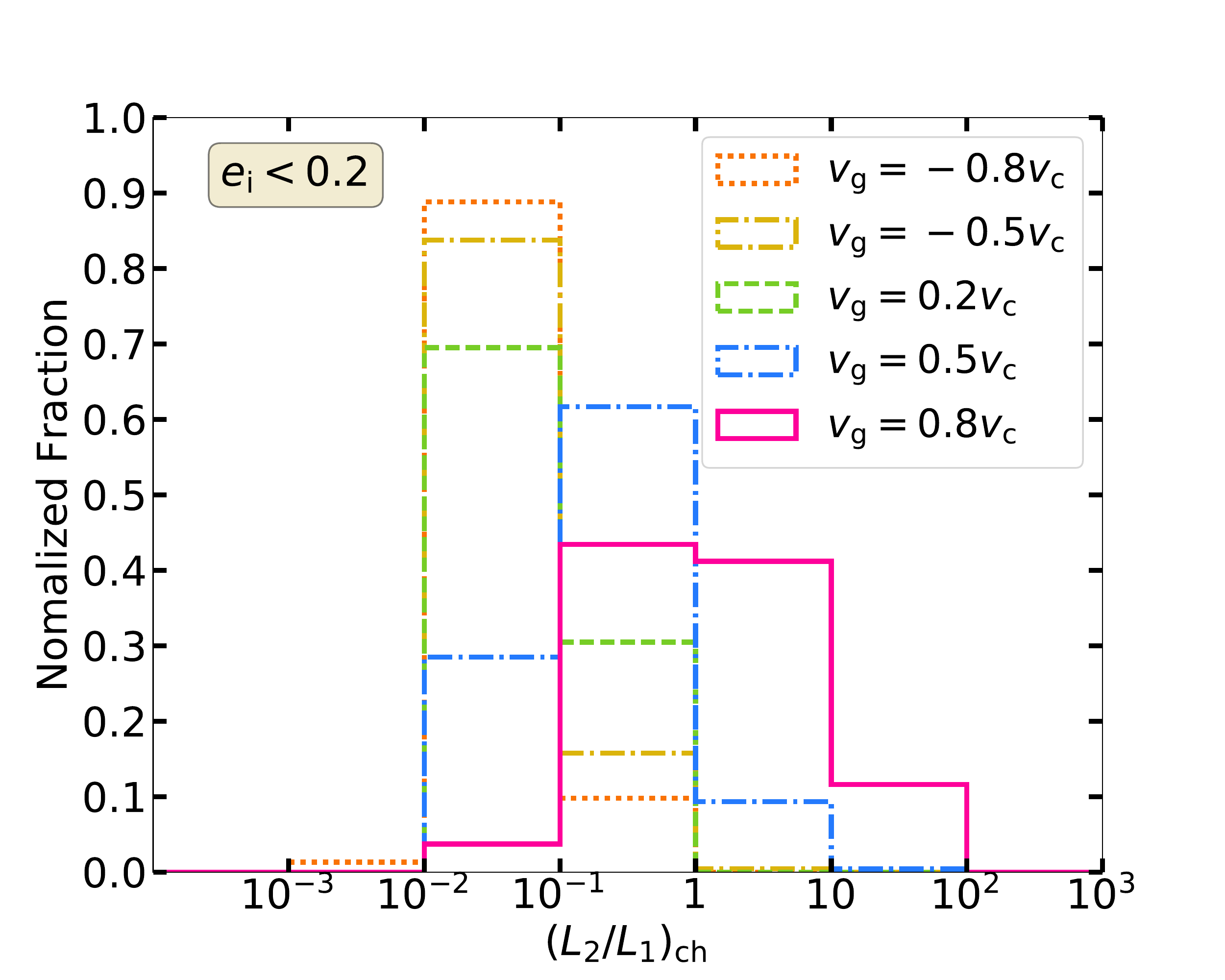}
            \includegraphics[width=0.49\textwidth]{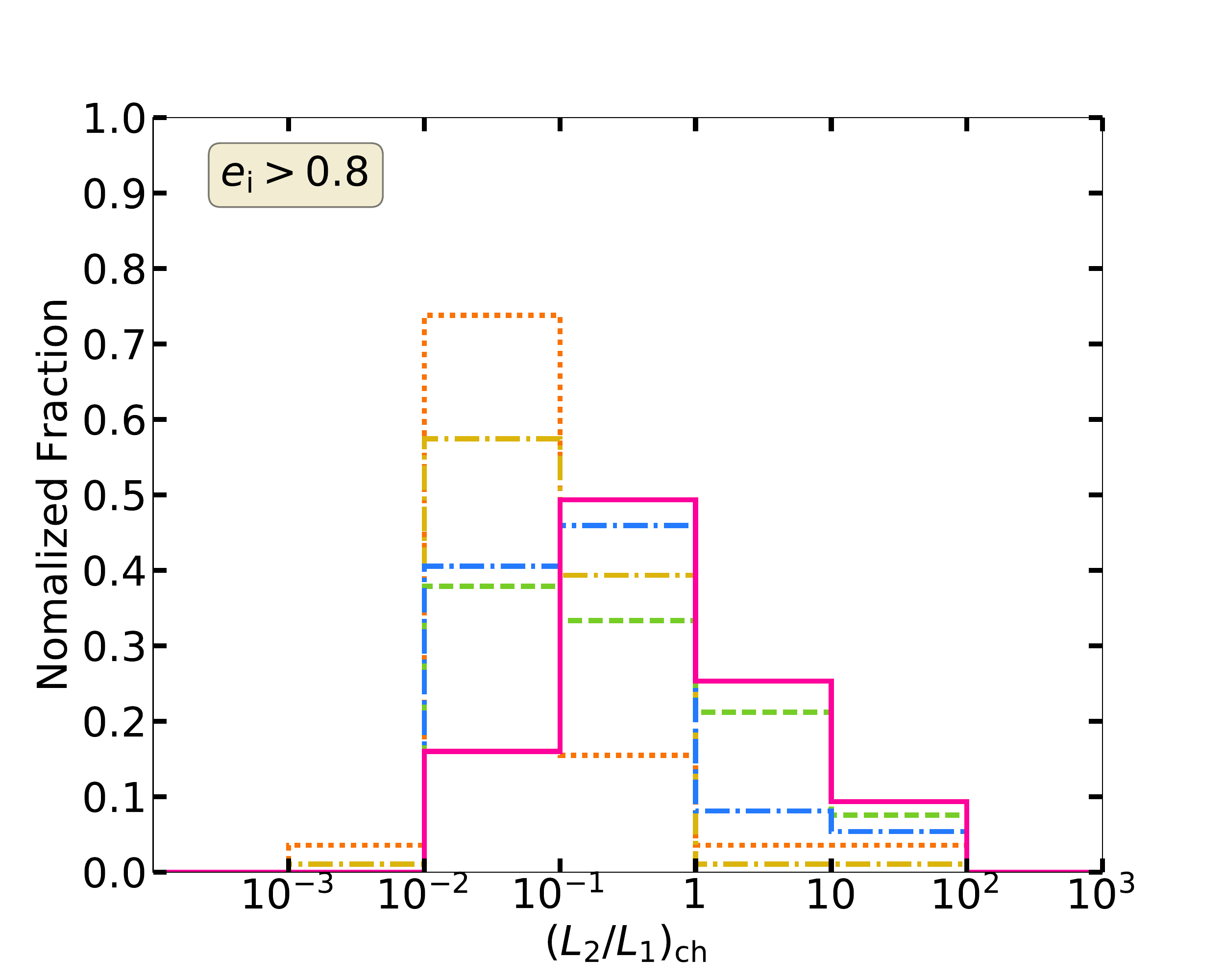}
\caption{Histograms of the characteristic dAGN luminosity ratios, \lmax, for \Ms\ on orbits with low (left) and high (right) initial eccentricity. Each panel shows the distribution of \lmax\ for different rotation speeds of the galactic disk. Positive (negative) values of \vg\ indicate sMBHs on prograde (retrograde) orbits.}
\label{fig:l_vg}
\end{figure*}
%
\begin{figure*}[t]
            \includegraphics[width=0.49\textwidth]{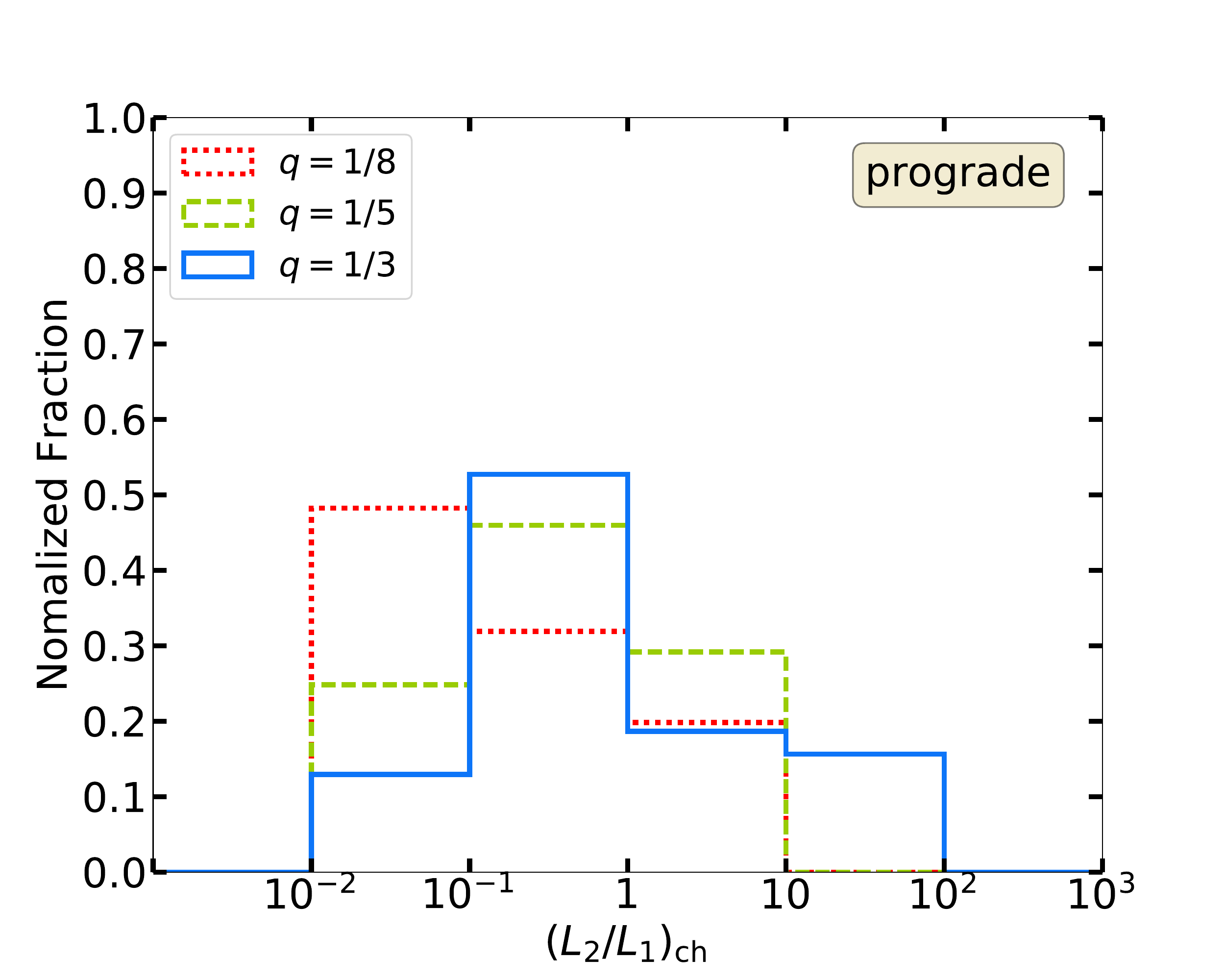}
            \includegraphics[width=0.49\textwidth]{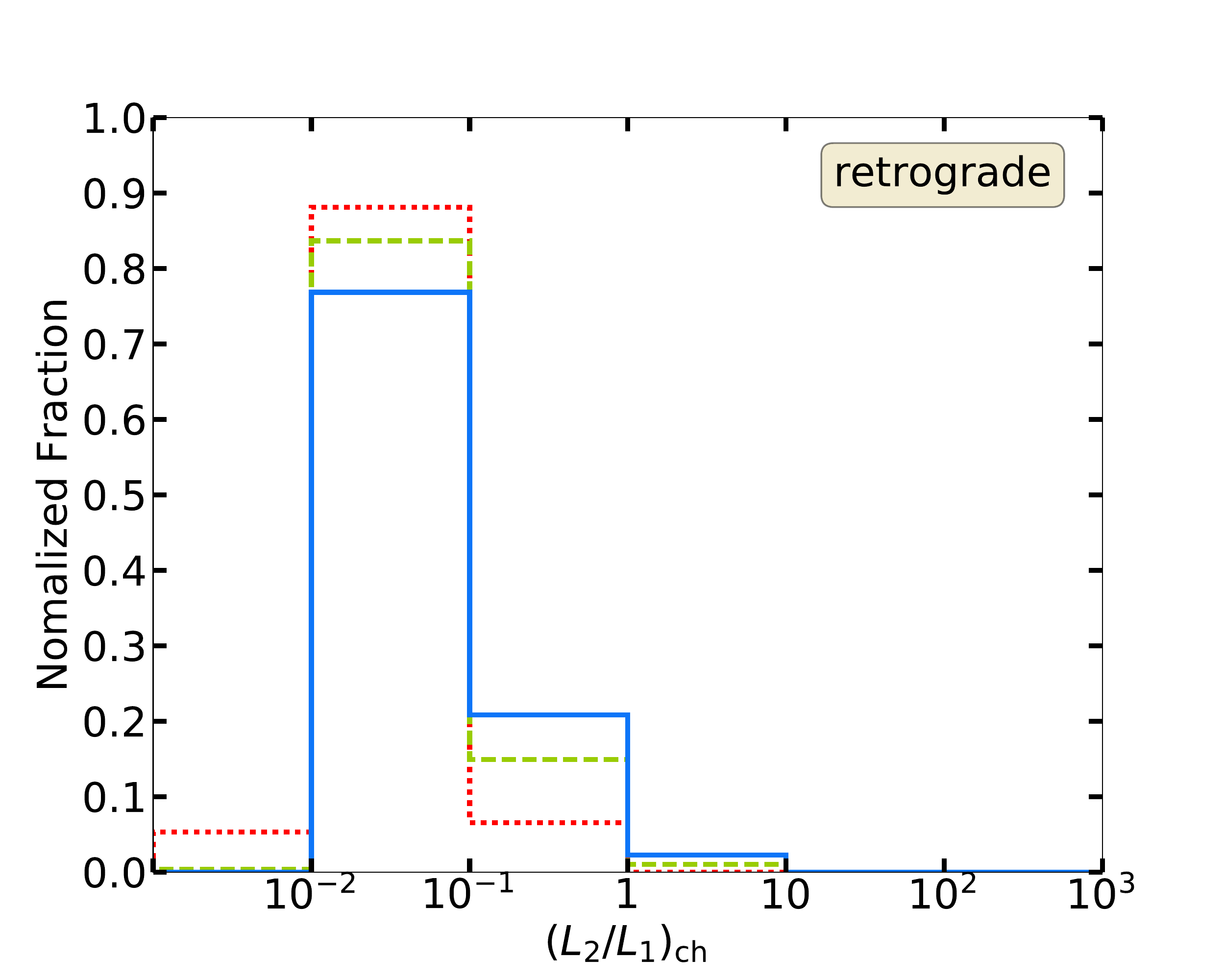}
\caption{Histograms of the characteristic dAGN luminosity ratios, \lmax, for \Ms s on prograde (left) and retrograde (right) orbits and different values of the mass ratio, $q$.
}
\label{fig:l_q}
\end{figure*}

The effects of the rotational speed of the galaxy on \lmax\ can be seen in
Figure~\ref{fig:l_vg}. The left panel shows that \Ms s on orbits with low initial eccentricity have larger \lmax\ values when the MBHs are co-rotating with the gas disks. This is particularly true as \vg\ tends to \vc. For example, nearly $80\%$ of dAGNs in gas disks with $v_{\rm g} = 0.8 v_{\rm c}$ have \lmax\ in the range of
$\sim 10^{-1} - 10$. On the other hand, $75\%$ of dAGNs where the \Ms\ is
on a retrograde orbit in a gas disk with $v_{\rm g} = -0.8 v_{\rm c}$
have \lmax\ in the range of $10^{-2}\sim 10^{-1}$. This effect is due
to the magnitude of the relative velocity, since MBHs on
circular prograde orbits have a smaller $\Delta v$, which results in a
larger accretion rate onto the \Ms\ and a bigger \lmax. In contrast, a \Ms\ on a retrograde orbit
will have a large $\Delta v$, supressing the accretion rate and reducing
\lmax\ by orders of magnitude. Interestingly, the right panel 
of Figure~\ref{fig:l_vg} shows that $\Delta v$ has a weaker effect when 
$e_i > 0.8$. In these cases, the \Ms s spend
significant time in the outer (low density) regions of the gas disk, and are
therefore most likely to be observed with a \lmax$< 1$, regardless of \vg.

The mass ratio of the two MBHs also impacts the distribution of
\lmax, as shown in Figure~\ref{fig:l_q}. For prograde orbits, dAGNs with larger \q\ have larger \lmax. For example, almost $60\%$ of dAGNs with $q=1/3$ have \lmax\ in the range of $10^{-1} \sim
1$, while nearly $70\%$ of those with $q=1/8$ have \lmax\ in the range
of $10^{-2} \sim 10^{-1}$. This is because the Bondi accretion rates on both MBHs are proportional to the square of their mass and thus, a larger mass ratio leads to a
larger luminosity ratio. The right panel of Figure~\ref{fig:l_q} shows that \lmax\ distribution of
dAGNs where the \Ms s are on retrograde orbits are not strongly impacted by an increase in $q$. In such  cases, the relative velocity between the \Ms\ and the gas disk is always large enough to counteract the
increase in luminosity ratio caused by a higher mass ratio.

\begin{figure*}[t]
\centering
     \includegraphics[width=\textwidth]{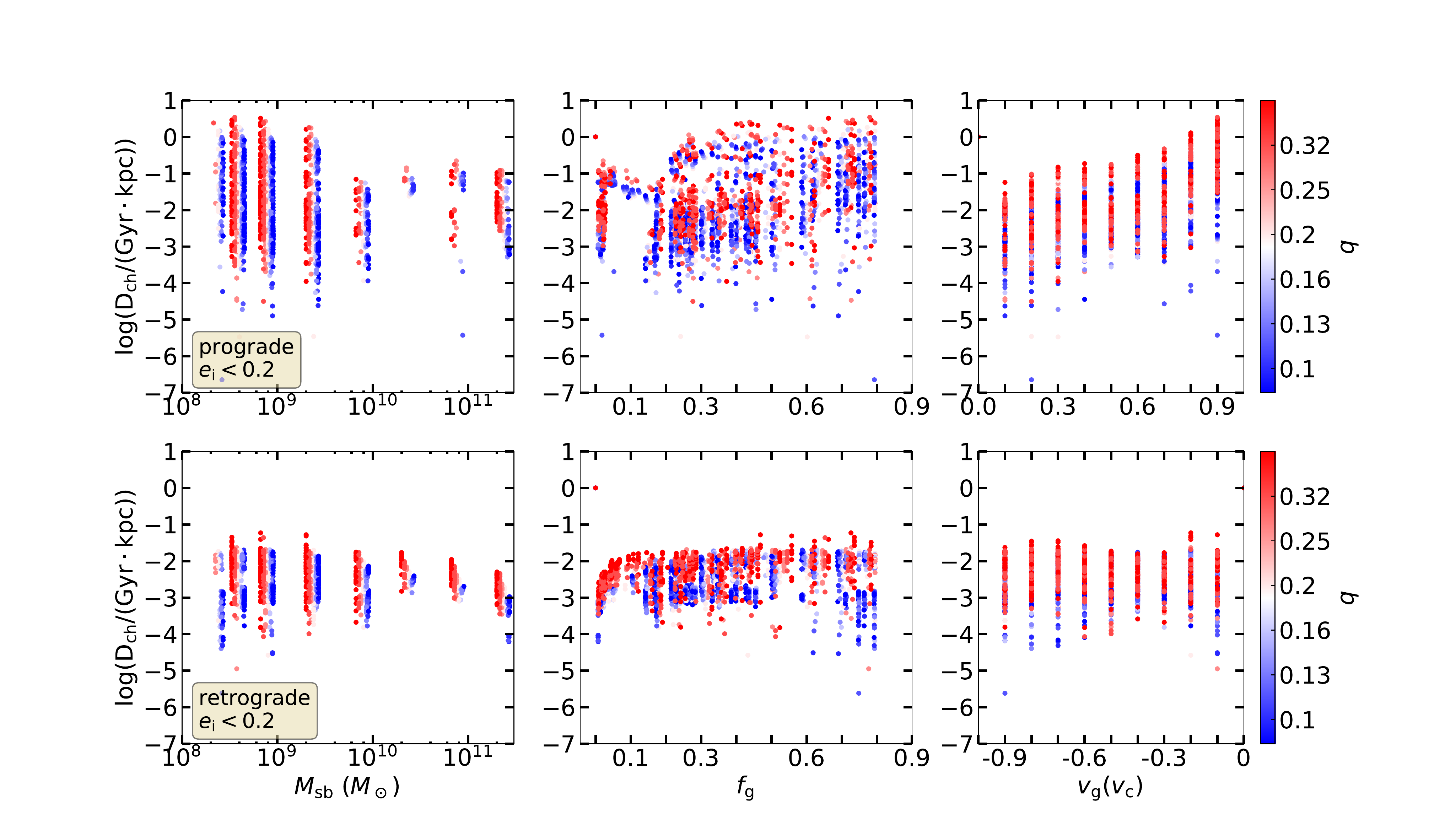}
\caption{The detectability, \D, of kpc-scale dAGNs  with $e_i < 0.2$ plotted as a function of the stellar bulge mass of  the merger remnant galaxy ($M_{\rm sb}$; left panels), the total gas fraction of the galaxy within 1~kpc ($f_g$; middle) and the rotation speed of the galactic disk (\vg; right). The top (bottom) panels correspond to \Ms s on prograde (retrograde) orbits. Each dot corresponds to one merger model in our suite and the color denotes the mass ratio of the pair, $q$. The most detectable dAGNs have \Ms s on preferentially prograde orbits and are found in gas dominated galaxies with rapidly rotating disks.} 
\label{fig:detect_el}
\end{figure*}

In summary, our simulations show that kpc-scale dAGNs are most likely to be
observed with \lmax$< 1$ or even $< 0.1$.  More favorable
luminosity ratios are however possible if the \Ms\ is on a prograde, low
eccentricity orbit, particularly in a rapidly rotating, high mass
galaxy. A larger $q$ is also expected to increase the characteristic luminosity ratio
of such dAGNs.

\subsection{The Detectability of kpc-scale Dual AGNs}
\label{sub:detectability}

Of all systems considered in previous sections, we expect configurations that lead to long lived, widely
separated dAGNs with high luminosity ratios to be more likely to be
detected. The combination of these three properties (separation, luminosity ratio,
evolution timescale) therefore determines the detectability of a dAGN. As all of
these properties are sensitive to the orbital and galactic properties,
it is important to to understand what types of galaxies and orbits
will lead MBH pairs that are preferentially
detected as dAGNs.
To address this question, we calculate the detectability, \D, for all dAGN models in our suite:
\begin{equation}
  \label{eq:D}
  D_{\mathrm{ch}} = d_{\mathrm{ch}} \times (L_2/L_1)_{\mathrm{ch}} \times f_t \left[ d_{\mathrm{ch}}, (L_2/L_1)_{\mathrm{ch}} \right] \times t_{\mathrm{evol}}.
\end{equation}
That is, \D\ for is the product of the most probable separation (in kpc) and luminosity
ratio and separation, the fraction of the total evolution time it spends at those two values, and the total evolution time of the MBH pair (in Gyr). For any given combination of $(d_{\mathrm{ch}}, (L_2/L_1)_{\mathrm{ch}},  t_{\mathrm{evol}})$, a kpc-scale dAGN is more likely to detectable if it has a larger value of \D. 

\begin{figure*}[t]
\centering
     \includegraphics[width=\textwidth]{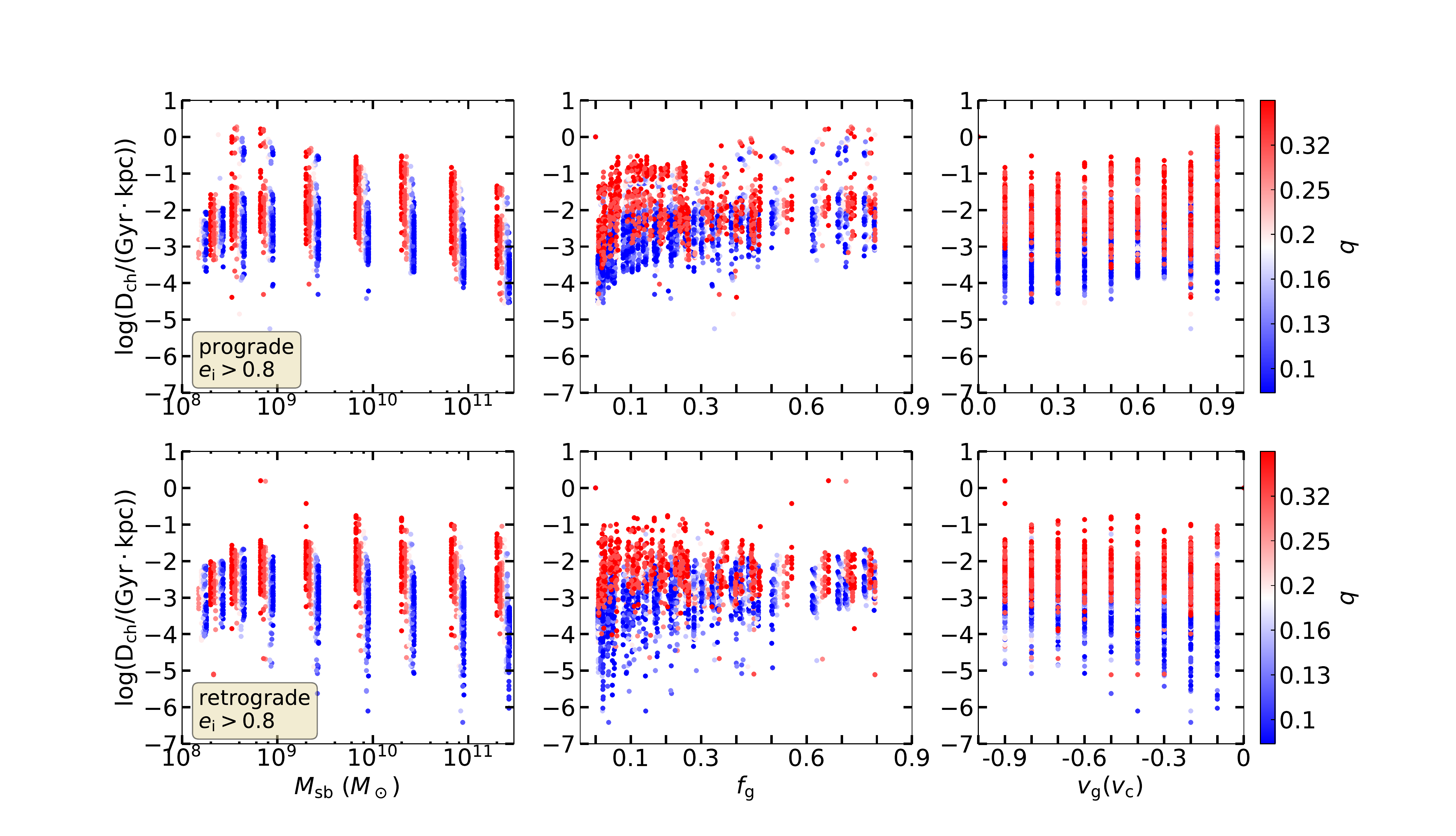}
\caption{As Figure~\ref{fig:detect_el}, but now showing the
  detectability \D\ for dAGNs where the \Ms\ is on a high initial eccentricity orbit.}
\label{fig:detect_eh}
\end{figure*}

Figures~\ref{fig:detect_el} and~\ref{fig:detect_eh} show \D\ for models with low and high initial orbital eccentricity, respectively, and the sub-panels illustrate the dependence on the key properties of the merger remnant galaxy. We define the total gas fraction of the model galaxy within 1~kpc, shown as a parameter in the middle panels of both figures, as
\begin{equation}
  \label{eq:fgas}
  f_g=\frac{M_{\rm gd,1}}{(M_{\rm gd,1}+M_{\mathrm{\star}})}\,,
\end{equation}
where $M_{\mathrm{\star}}$ is the total mass of the bulge and
stellar disk within this radius.  


Many of the results found in the previous section can also be identified in
Figure~\ref{fig:detect_el}. In general, dAGNs with a \Ms\ on a $e_i <
0.2$, prograde orbit are more detectable than retrograde ones, because the sMBHs on prograde orbits
are characterized by higher luminosity ratios and larger \D\ (e.g., Figure~\ref{fig:l_vg}). 
Considering the prograde dAGNs more closely, we find that those in host galaxies with small $M_{\rm sb}$,
large \fg, \vg\ close to \vc, and large $q$ are the most
detectable. As discussed in Sect.~\ref{sub:lratio}, dAGNs
with large $q$ or dAGN in galaxies with
$v_{\rm g} \approx v_{\rm c}$ are more likely to produce a higher $L_2/L_1$ and are
therefore more easily detectable. The dAGNs in host galaxies with $M_{\rm sb} <
10^{10}$~M$_{\odot}$ or $f_g > 0.2$ are more detectable because the
orbital decay in these galaxies is frequently dominated by gaseous
DF. LBB20a showed that in these cases the \Ms\ has a longer
evolution time compared to those systems where the DF due to the stellar bulge dominates the orbital evolution. The longer timescale for orbital decay means that sMBHs in these galaxies
spend more time at larger separations, and are therefore more easily 
detectable as dAGNs.

The detectability of dAGNs with \Ms s that have $e_i > 0.8$ are shown
in Figure~\ref{fig:detect_eh}. The plots indicate that eccentric and prograde dAGNs are on average slightly less detectable than their counterparts with $e_i < 0.2$. The apocenters of $e_i > 0.8$ orbits are located further out in the galactic gas disk, where the gas density is low. This makes the accretion rate of the
sMBHs, the luminosity ratio, and the \D\ lower compared to those of dAGNs in host galaxies with the same properties but of low initial eccentricity.

Figure~\ref{fig:detect_eh} also shows that when it comes to systems with $e_i > 0.8$ prograde orbits, their detectability peaks in a narrow range of bulge masses centered on $M_{\rm sb}
\approx 5\times 10^{8}$~M$_{\odot}$. The initial rise in \D\ with increasing $M_{\rm sb}$ can be understood because the accretion rate of sMBHs and the luminosity ratio are larger in more massive galaxies (see Sect.~\ref{sub:lratio}). However, as $M_{\rm sb}$ increases the evolution time of dAGNs shortens, which causes \D\ to reach a maximum at  $M_{\rm sb} \approx 5\times 10^{8}$~M$_{\odot}$ and then decrease. This is a reflection of the fact that in larger stellar bulges orbital evolution of the \Ms\ becomes increasingly dominated by stellar DF, and therefore faster, on average.

\begin{figure*}[t]
\centering
     \includegraphics[width=\textwidth]{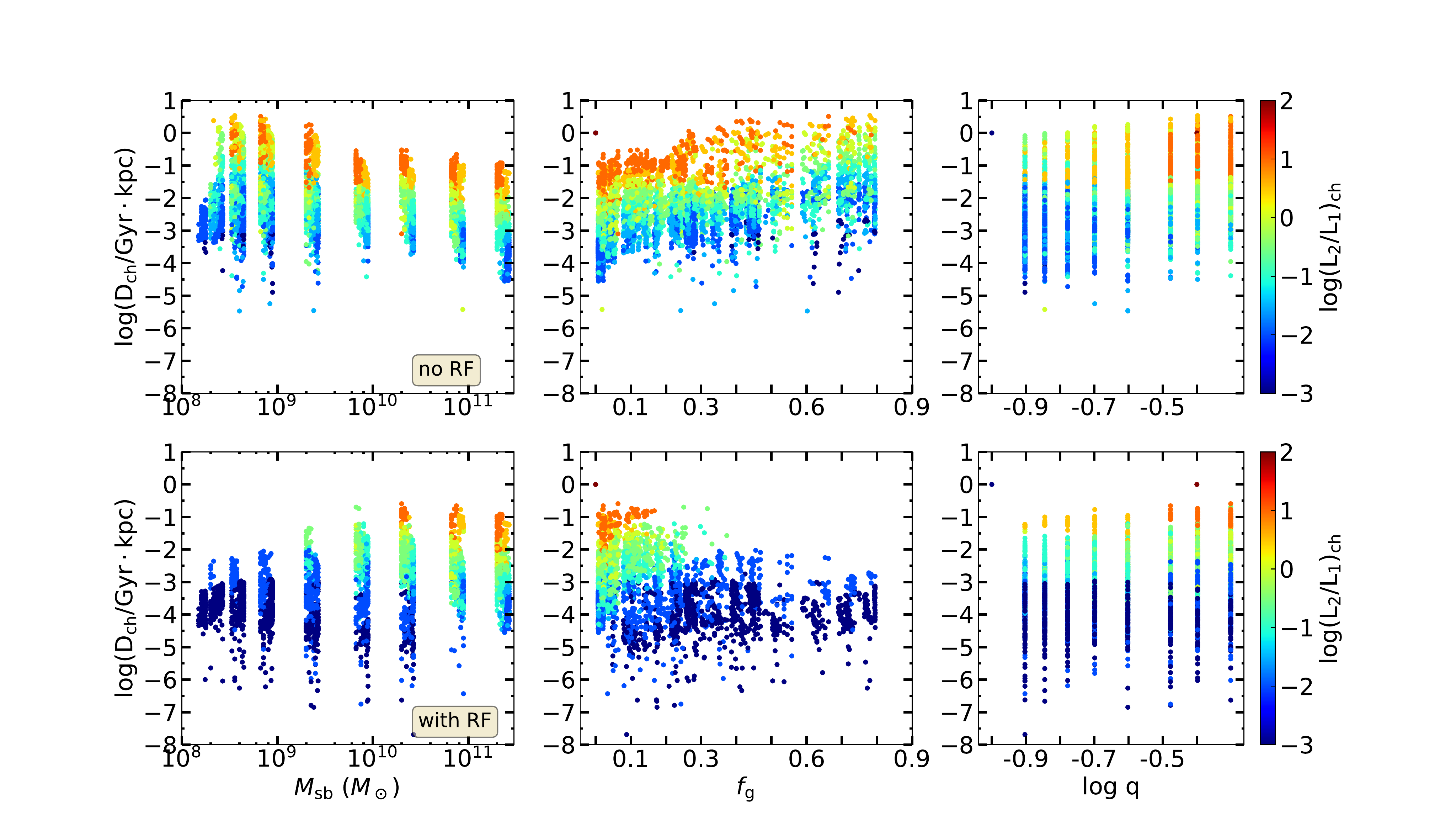}
\caption{Detectability of dAGNs in the models without (top) or with (bottom) the radiation
feedback effects. Dots represent dAGN systems on prograde orbits regardless of eccentricity and the color bar marks \lmax.}
\label{fig:compare}
\end{figure*}

Dual AGNs on retrograde $e_{\rm i} > 0.8$ orbits are more easily detectable than those on retrograde $e_{\rm i} < 0.2$ orbits (see bottom rows in Figure~\ref{fig:detect_eh} and Figure~\ref{fig:detect_el}). This is
because the orbital velocity of a sMBH on a more eccentric retrograde orbit is smaller at the apocenter, which makes the $\Delta v$ smaller as well. The lower relative speed increases the efficiency of BHL accretion onto the \Ms\ and consequently, $L_2/L_1$.

In summary, the results shown in Figure~\ref{fig:detect_el}
and~\ref{fig:detect_eh} indicate that dAGNs are more likely to be
detected in gas-rich post-merger galaxies with rapidly
rotating disks. In addition, the detectability is increased for large
$q$ and a \Ms\ that is on a low eccentricity prograde orbit. In
contrast, dAGNs where the \Ms\ is on a $e_{\rm i}<0.2 $, retrograde
orbit are the least detectable systems in any post-merger galaxy remnant. 

\subsection{The Impact of Radiation Feedback on dAGN Detectability}
\label{sub:RF}

\begin{figure*}[ht]
\centering
     \includegraphics[width=0.9\textwidth]{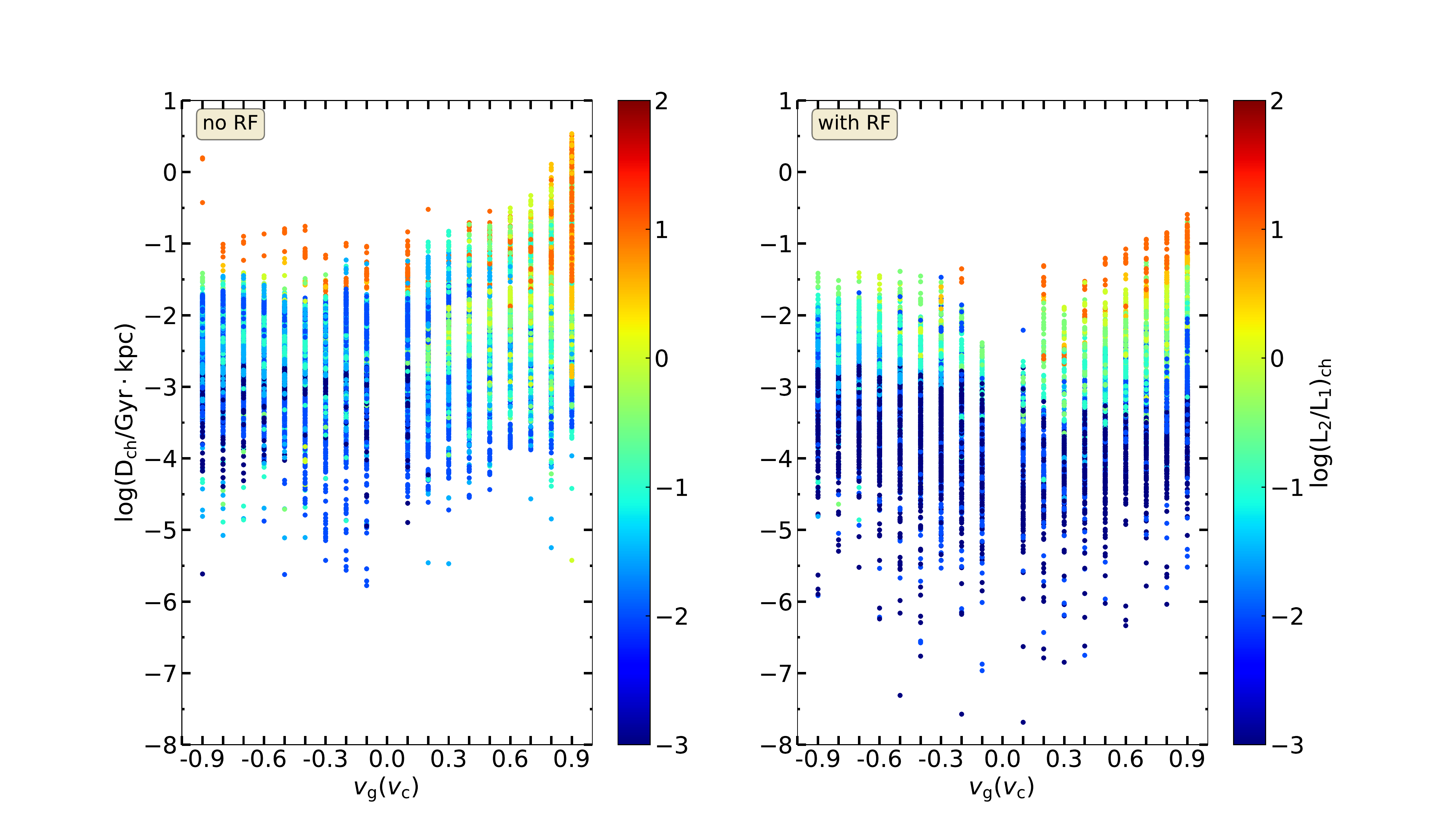}
\caption{Detectability of dAGNs as a function of the rotation speed of the galaxy remnant in models without (left) or with (right) the radiation feedback effects. Negative (positive) values of \vg\ represent \Ms s on retrograde (prograde) orbits. The color bar marks \lmax. }
\label{fig:compare_vg}
\end{figure*}

The radiation produced by each MBH of a dAGN can influence
both the luminosity and dynamics of the system. The
thermal pressure of the ionized bubble surrounding an accreting
MBH regulates its accretion rate \citep[e.g.][]{Ostriker1976,
  B1985, Ric2008, Park2011, Park2012}, potentially suppressing the
  emitted luminosity. The magnitude of this effect depends on the
  motion of the MBH, as well as the density and temperature of the
  surrounding gas \citep{Park2013}. Similarly, the
  shape of the ionized bubble can affect the DF force for a MBH
  moving in a gas-rich medium \citep{PB2017, G2020,T2020}. 
  Below, we investigate how these phenomena impact the predicted detectability of dAGNs.

We first consider the accretion rates and resulting luminosities of the stationary \Mp\ and the moving \Ms\ in the presence of radiative feedback. To calculate the accretion rate onto the \Mp\ in the presence of radiative feedback we use the parametric model developed by \citet{Park2012}, calibrated on their radiation hydrodynamic simulations. Specifically, we make use of their equation~(7), (16) and (21), with a spectral index of 1.5 and radiative efficiency of 0.1. We find that the radiation feedback does not affect the accretion rate and luminosity of most pMBHs in our models, because the gas densities surrounding them are sufficiently high to counter the effect of the radiation pressure. This places most pMBHs in our model in the so-called hyper-Eddington accretion regime \citep{I2016}, characterized by mass accretion rates larger than the Eddington rate and emergent luminosities $\lesssim L_{\rm Edd}$, limited by photon trapping in the high density gas. Hence, we limit the luminosity of the pMBHs in this regime to 10\% of the Eddington luminosity.

The moving \Ms\ on the other hand finds itself in the regime where radiation feedback can strongly suppress its accretion rate (see Appendix A). Here, we adopt the parametric model from \citet{Park2013} to calculate the reduced accretion rate onto the \Ms\ in the presence of radiative feedback. They showed that  the accretion rate onto a moving MBH increases as $\propto \Delta v^2$ until $\Delta v \approx 2c_{\rm s, in}$ (where $c_{\rm s, in} \approx 2 c_{\rm s,\infty}$ is the sound speed inside the ionized region), and beyond this value asymptotes to the classical BHL solution:
\begin{equation}
\label{eq:RF_M2}
\dot{M}_{\rm 2} \approx \frac{\rho_{\rm \infty} (G\;M_{\rm 2})^2}{c_{\rm s, in}^3}\times
  	\begin{cases}
     0.7\;(\frac{\Delta v}{2\; c_{\rm s, in}})^2, \; 0<\Delta v \leq 2\; c_{\rm s, in}\\
    (1+(\frac{\Delta v}{c_{\rm s, in}})^2)^{\rm -3/2}, \;\; \Delta v \gg 2\; c_{\rm s, in}.
    \end{cases}
\end{equation}
The first line of equation~(\ref{eq:RF_M2}) indicates that the
accretion rate of the \Ms\ is significantly suppressed relative to the
BHL rate when the Mach number $\mathcal{M}=\Delta v/c_{\mathrm{s,}
  \infty} < 5$ (where the gas temperature in the regions unaffected by
the \Ms\ is calculated as in Sect.~\ref{sub:DF}). We therefore expect the radiative feedback effects to be more prominent for sMBHs on
prograde orbits, when $\Delta v$ is relatively small and the Mach number remains in this range.

Radiation feedback not only affects the accretion rate but may also change how DF influences the inspiral of the \Ms\ -- we consider this effect next. Depending on the motion of the MBH relative to the background medium, as well as the properties of the ionized gas, the shape of the gaseous wake around the MBH can reverse the direction of the DF force and speed up the moving MBH \citep[an effect referred to as ``negative DF'';][]{PB2017,G2020,T2020}. LBB20b modeled this effect and studied how negative DF changes the evolution of MBH pairs in different types of host galaxies. They find that radiation feedback leads to longer \tevol, is more important in merger
remnants with gas fraction $f_{\rm g}>0.1$, rotation speeds $v_g \sim v_c$, and when sMBHs are on
prograde, low eccentricity orbits. We therefore expect radiative feedback by the \Ms\ to impact the detectability of some fraction of dAGNs due to these dynamical effects resulting in longer \tevol.

To quantify the effects of radiative feedback we reran our full dAGN model suite (described in Section~\ref{sec:methods}) with parametric models described in previous paragraphs. We evaluate the detectability of each new model and compare it to models in the absence of radiation feedback in Figure~\ref{fig:compare}. Since the effects of radiative feedback are strongest for sMBHs on prograde orbits, we only show \D\ for these type of systems. 

The colorbar of Figure~\ref{fig:compare} indicates dAGNs in models with radiative feedback show a significant drop in \lmax\ caused by the suppression of accretion onto the \Ms.  Although
the effects of negative DF typically increase \tevol\ (LBB20b), the
competing decrease in \lmax\ completely overtakes it, resulting in 
a lower overall value of \D\ for dAGNs affected by radiation. Since radiative feedback is more likely to arise in gas-rich environments, the suppression  of \D\ is most noticeable when the merger remnant galaxy is dominated by the gas disk (which in our models corresponds to $M_{\rm sb} < 10^{11} M_{\rm \odot}$ or $f_{\rm g}>0.2$). The drop in \D\ does not have a strong relation with $q$ however, meaning the effect of radiative feedback does not have a strong preference for the mass ratio of dual AGNs.

Because the \Ms\ accretion rate and luminosity in the models with
radiative feedback effects are a sensitive function of the relative
velocity between the \Ms\ and the rotating gas disk in the remnant galaxy (equation~(\ref{eq:RF_M2})), we also examine the dependence of detectability on this factor. The left panel of Figure~\ref{fig:compare_vg} illustrates that in the absence of the radiation feedback, the dAGNs in galaxies with $v_g \approx v_c$ are more easily detectable, because this condition leads to a larger BHL accretion
rate. The right panel of Figure~\ref{fig:compare_vg} shows that in the presence of feedback, the same group of systems has reduced detectability, due to the suppression of accretion onto the \Ms. This is because, as noted before, the drop in $L_{\rm 2}$ due to radiative feedback is most significant in
the range of $0<{\cal M}<5$, which here roughly corresponds to systems with $v_g > 0.5 v_c$.

For simplicity, Figure~\ref{fig:compare} and~\ref{fig:compare_vg} show the combined results for sMBHs on both low and high eccentricity orbits. It is worth noting however that the decrease in \D\ due to radiation feedback is about $\sim 100$ times more pronounced for low eccentricity orbits. This is because the systems with eccentric orbits exhibit lower \Ms\ accretion rates independent of the presence of radiation feedback effects, since they spend a significant portion of their time close to the orbital apocenter, in the region of the galaxy remnant where the gas density is low. Thus, their (low) detectability is not significantly affected by radiation feedback.

In summary, dAGNs in models with radiative feedback show a significant
drop in detectability caused by the suppression of accretion onto the
\Ms. The suppression of detectability is most significant when the
merger remnant galaxy is dominated by the gas disk ($M_{\rm sb} <
10^{11} M_{\rm \odot}$ or $f_{\rm g}>0.2$) and the \Ms\ is on a
prograde orbit within a rapidly rotating galaxy. On the other hand,
the effects of radiative feedback are laregely independent on the the
mass ratio of the MBH pair.

\section{Discussion}
\label{sec:discuss}

\subsection{Implications for Observations of Dual AGNs}
\label{sub:implication}

Observationally identifying sub-kpc dAGNs is a daunting challenge with a relatively low number of 
unambiguously confirmed cases \citep[on the order of $\sim 10$;][]{Rosa2020}. dAGN candidates are often selected
from catalogs of AGNs that exhibit double-peaked
[\ion{O}{3}] $\lambda$5007\AA\ emission lines. However, since there are
multiple plausible causes for existence of the double-peaked lines in the AGN spectra \citep[e.g.,][]{Crenshaw2010,Rosario2010,Muller2015}, follow-up
high-resolution imaging or spatially resolved spectroscopy is
needed to identify two distinct nuclear sources in the
candidate host galaxy \citep[e.g.,][]{C2012,Fu2012,Hou2019,Hou2020,Foord2020}. This is an observationally 
expensive strategy to identify dAGNs and the results in this work
can provide guidance on how to best prioritize follow-up observations of candidate dAGNs.

The most important property of the host galaxies that enhances the
detectability of dAGNs is the rotation speed, \vg. If the \Ms\ is on a
prograde orbit then it is more likely to spend a large fraction of its
orbit with a low $\Delta v$, boosting its luminosity (e.g.,
Figure~\ref{fig:l_vg},~\ref{fig:detect_el} and~\ref{fig:detect_eh}). Similarly, if
the eccentricity of the orbit remains moderately low, then the evolution of the
orbit can be slow enough that there is a good chance that the dAGN
will be observed at a separation of $\sim 0.7 - 0.8$~pc
(Figure~\ref{fig:d_vg}).

The size and relative mass of the gas disk in the candidate galaxy is
another important element in determining the detectability of
dAGNs. The orbital decay of the \Ms\ speeds up once it enters the
stellar bulge, so galaxies with either small bulges or massive gas
disks are be easier to detect as dAGNs, as the \Ms\ will be more likely
to be observed at larger separations. A larger gas fraction of the
galaxy should nominally also increase the detectability of dAGNs by enhancing the
likelihood of rapid accretion onto the \Ms. However, as shown in
Figure~\ref{fig:compare}, radiation feedback effects are also most important
in galaxies with large $f_g$, and could significantly decrease the
detectability by suppressing accretion onto the \Ms. If the radiation
feedback effects indeed act as described in Sect.~\ref{sub:RF}, then
large, relatively gas poor galaxies are the best candidates for
detecting dAGNs. In this case, the rapid evolution of the \Ms\ in the
large bulge is compensated by the overall increase in the characteristic separation of the
system (e.g., Figure~\ref{fig:d_m}).

Our results also indicate that identified sub-kpc dAGNs will
most likely have large $q$ with a \Ms\ on a prograde orbit. As shown in Figure~\ref{fig:detect_el} and Figure~\ref{fig:detect_eh}, significantly more low $e_i$
models yield high \D\ than high $e_i$ ones. Therefore, it is likely that detected dAGNs will contain sMBHs on low eccentricity orbits.

\begin{figure*}[t]
            \includegraphics[width=0.49\textwidth]{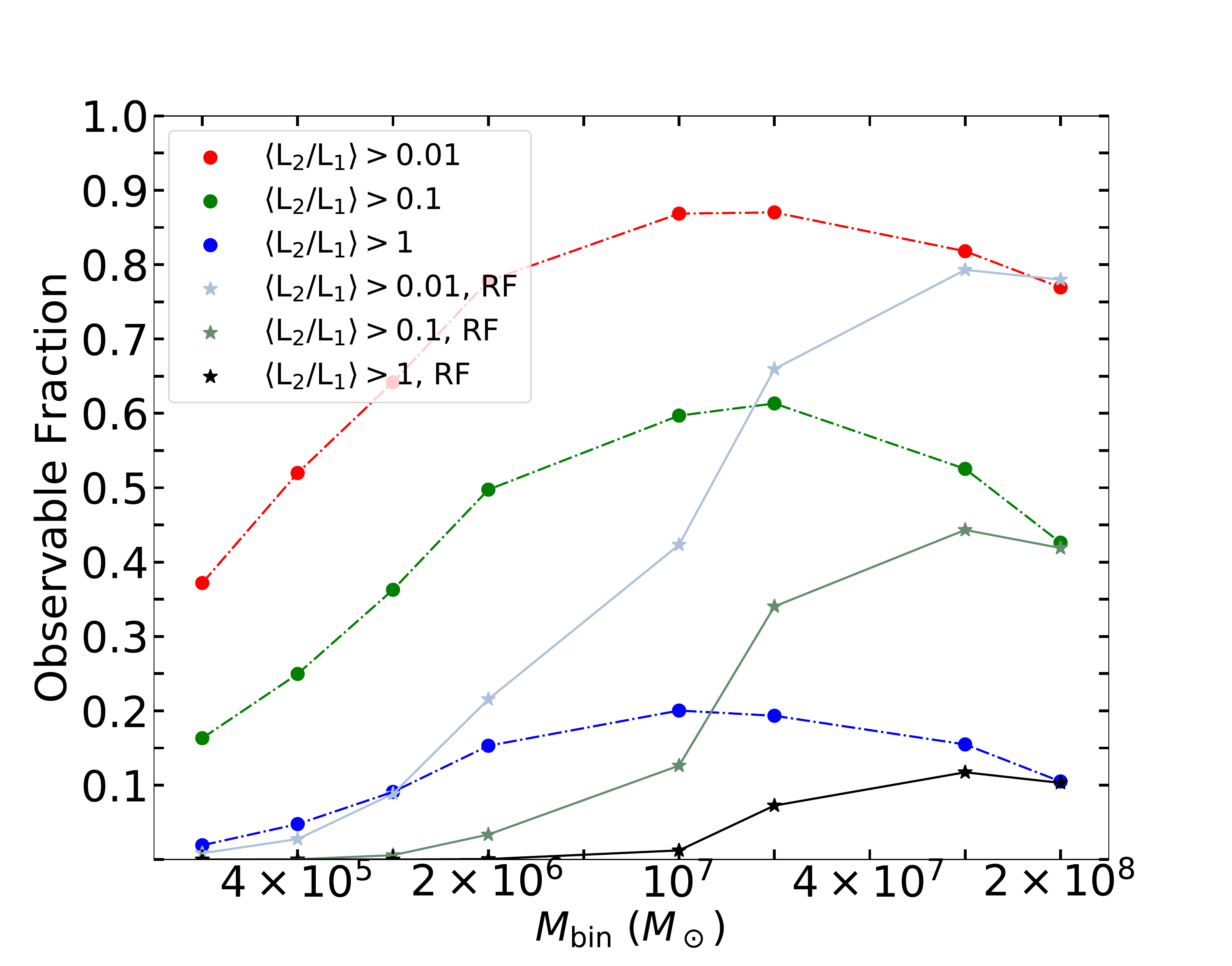}
            \includegraphics[width=0.49\textwidth]{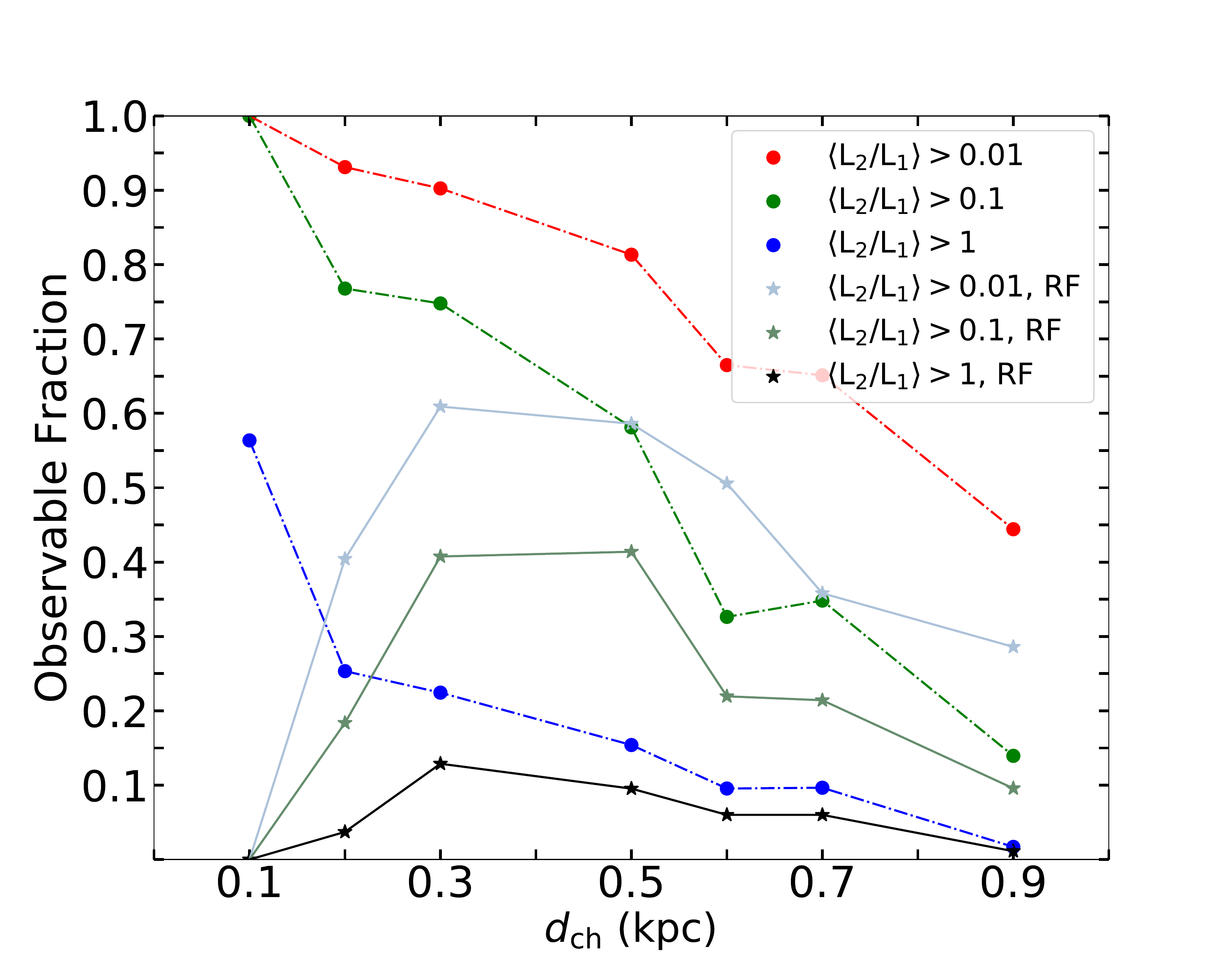}
\caption{Fraction of dAGNs that can be observed above time averaged luminosity ratio ($\langle L_{\rm 2}/ L_{\rm 1} \rangle$) of $0.01, 0.1, 1 $ at different pair mass (left) and characteristic separation (right).}
\label{fig:detect_ratio}
\end{figure*}

Another demonstration of the challenges and tradeoffs of observationally searching for dAGNs is shown in Figure~\ref{fig:detect_ratio}. Figure~\ref{fig:detect_ratio} illustrates the observable fraction of dAGNs which is defined as the fraction of dAGNs whose time averaged luminosity ratio ($\langle L_{\rm 2}/ L_{\rm 1} \rangle$) through out the evolution process is larger than a threshold value. In the absence of radiation feedback, dAGNs of $M_{\rm bin} \sim 10^{\rm 7} M_{\rm \odot}$ have the highest observable fraction in our simulations (left panel of Figure~\ref{fig:detect_ratio}). Nearly $90 \%$ of these dAGNs are observable if setting $\langle L_{\rm 2}/ L_{\rm 1} \rangle > 0.01$ to be the limit. As the pair mass increases, the observable fraction of dAGNs first increases then turns over. The rise is due to the increasing disk scale radius, which leads to a slower gas density drop and results in a higher mean luminosity ratio. While the turn over is because of the fast evolution rate inside the bulge which reduces the time these dAGNs shine with high luminosity ratio. However, including radiation feedback drops the observable fraction to $40 \%$ assuming the same limit. On the other hand, the observable fraction of dAGNs with $M_{\rm bin} \sim 2 \times 10^{\rm 8} M_{\rm \odot}$ is larger than $75 \%$ and only weakly affected by radiation feedback effects since these models have very large bulges and low gas fractions which make the gas based radiation feedback effect negligible. Thus, we expect dAGNs with higher pair mass to have a higher observable fraction in observational surveys of dAGNs.

Almost all dAGNs with characteristic separations smaller than $0.3$ kpc have $\langle L_{\rm 2}/ L_{\rm 1} \rangle$ above $0.01$ (right panel of Figure~\ref{fig:detect_ratio}). The curves rise to low separation because as the separation becomes smaller, the gas density becomes larger increasing both the luminosity ratio and the observable fraction. However, the suppression of accretion onto the sMBH due to radiation feedback is predicted to significantly suppress the observable fraction at small separations. Thus, after accounting for the impact of radiation feedback, separations of $\sim 0.3-0.5$ kpc will lead to the highest observable fractions of dAGNs.


\subsection{Impact of Simplifying Assumptions}
\label{sub:assumptions}

The advantage of our semi-analytic model is the ability to run
simulations over a wide range of galaxy and orbital properties at
the cost of making some simplifying assumptions. The potential impact
of our assumptions on the dynamical aspects of evolution of MBH pairs is discussed
in LBB20a and LBB20b. Here, we consider the possible effects of these
assumptions on the derived dAGN properties.

We assume the \Mp\ is fixed at the center of the host galaxy. If the
motion of the \Mp\ and its orbital decay due to DF forces are included
in the simulations, the resulting \tevol\ would be shorter, thus reducing
\D. This effect would be strongest in comparable mass MBH pairs and weaker in those with small
$q$. However, if the \Mp\ is moving, it would accrete at the BHL rate, which is smaller than the Bondi rate used in the current model, increasing $L_2/L_1$ and the detectability. Again, this
effect would be stronger for comparable mass MBH pairs, since $\Delta v$ of the \Mp\ would be maximized in this case. As the these two effects compete against each other, we do not expect a large systematic discrepancy in \D\ due to this assumption.

Similarly, the orbit of the \Ms\ is assumed to always reside in the
midplane of the galaxy. If inclined orbits take the \Ms\ outside of the gas disk, \tevol\ will increase since the gaseous DF force would be less efficient. For highly inclined orbits however, the \Ms\ will spend a large
fraction of time in the low gas density regions and the characteristic
luminosity ratio, \lmax, would likely be many orders of
magnitude smaller relative to a \Ms\ on a co-planar orbit. This effect is
likely to overtake the increase in \tevol\ and we therefore expect the
detectability of dAGNs where the \Ms\ is on an inclined orbit to be
much smaller than for the co-planar systems.

The masses of the two MBHs in the models are held fixed during our calculations. If we allow the MBHs to gain mass, we find that the average change in $q$ is $\lesssim 60 \%$. By assuming a fixed mass ratio throughout the evolution, we provide a lower limit of the detectability. If the growth of mass of MBHs by accretion is taken into account, we expect the detectability to increase by less than an order of magnitude independent of the presence of radiation feedback. In the context of Figure~8 $\&$ 9, we expect the distribution shifts towards larger $M_{\rm tot}$, smaller $f_{\rm g}$, and larger detectability due to the increased mass ratio in the evolution. 

\section{Conclusions}
\label{sec:concl}

We present the results of nearly 40,000 simulations of model dAGN systems, in which we follow the orbital evolution of the \Ms\ as it decays in response to the DF forces from the gas and stellar components of the merger remnant galaxy. For each dAGN, we calculate how orbital separation and luminosity ratio of the two MBHs change as a function of the properties of the galaxy and sMBH orbit and use them to evaluate the most probable separations, \dmax, and luminosity ratios, \lmax. Together with the the evolution timescale, \tevol, we use these properties to define the detectability and gauge which systems are most likely to be discovered as dAGNs in observations. We find that:
\begin{itemize}
\item The low, unequal mass MBH pairs (e.g., $M_{\rm bin} = 3\times10^6\,M_{\rm \odot}$, $q=1/9$) in slowly rotating galaxies have most probable dAGN luminosity ratios of $\sim 10^{-2}$ and the most
probable dAGN separations $d \sim 0.3-0.4$\,kpc for prograde and $\sim 0.4-0.9$\,kpc for retrograde orbits (see \S~\ref{sub:prob}). 
\item The high, comparable mass MBH pairs (e.g., $M_{\rm bin} = 10^8\,M_{\rm \odot}$, $q=1/3$) in rapidly rotating galaxies on the other hand exhibit higher characteristic dAGN luminosity ratios and separations. The latter is a consequence of a relatively slow evolution of these systems at large separations, where they spend a significant fraction of their overall evolution time (\S~\ref{sub:prob}).
\item  The most likely dAGN separations can be as low as $\sim 0.1$\,kpc in prograde systems (especially in slowly rotating, low mass galaxies), whereas the characteristic dAGN separations for retrograde systems are $\sim 0.8$\,kpc for a wide range of remnant galaxy properties (\S~\ref{sec:d}).
\item The kpc-scale dAGNs are most likely to be observed with \lmax$< 1$ or even $< 0.1$. Of these, systems with larger MBH mass ratios in rapidly rotating, high mass galaxies tend to occupy the higher end of the luminosity ratio distribution (\S~\ref{sub:lratio}).
\item Overall, dAGNs are more likely to be detected in gas-rich post-merger galaxies with rapidly rotating disks. In addition, the detectability is increased in systems with comparable MBH mass ratios characterized by low eccentricity, prograde orbits. In contrast, dAGNs on retrograde, low eccentricity  orbits are some of the least detectable systems among our models (\S~\ref{sub:detectability}).
\item The above findings are formulated for systems in which the accreting MBHs do not exhibit radiative feedback. dAGNs in models with radiative feedback show a significant drop in \lmax\ caused by the suppression of accretion onto the \Ms, resulting in a lower overall detectability. Since radiation feedback is more likely to arise in gas-rich environments, the suppression in detectability is most pronounced in merger remnant galaxies dominated by the gas disk. 
Hence, if the radiation feedback effects operate as described here, then large, relatively gas poor galaxies are the best candidates for detecting dAGNs (\S~\ref{sub:RF}). 
\end{itemize}

In the next decade, new X-ray observatories (e.g., \textit{eROSITA},
\citealt{erosita}; \textit{Athena}, \citealt{athena}), radio surveys
(e.g., ngVLA, \citealt{ngvla}; SKA; \citealt{ska}), and optical
surveys (e.g., \textit{JWST}, \citealt{jwst}) will dramatically increase the
population of known dAGNs, especially at separations $\lesssim$\,kpc. The detectabilities computed from our models provide a convienent way to select dAGN candidates for follow-up
observations. In the future, comparing a sample of kpc-scale dAGNs luminosities
and separations to these results will provide a test of the nature and
efficiency of dynamical friction forces in transporting MBHs in
post-merger galaxies. 

\acknowledgments

T.B. acknowledges the support by the National Aeronautics and Space Administration (NASA) under award No. 80NSSC19K0319 and by the National Science Foundation (NSF) under award No. 1908042. We also acknowledge our anonymous referee for helpful comments.

\appendix
\section{The Distribution of Time-Averaged Accretion Rates onto the MBHs}
\label{m_dot}

\begin{figure*}[t]
\centering
     \includegraphics[width=0.5\textwidth]{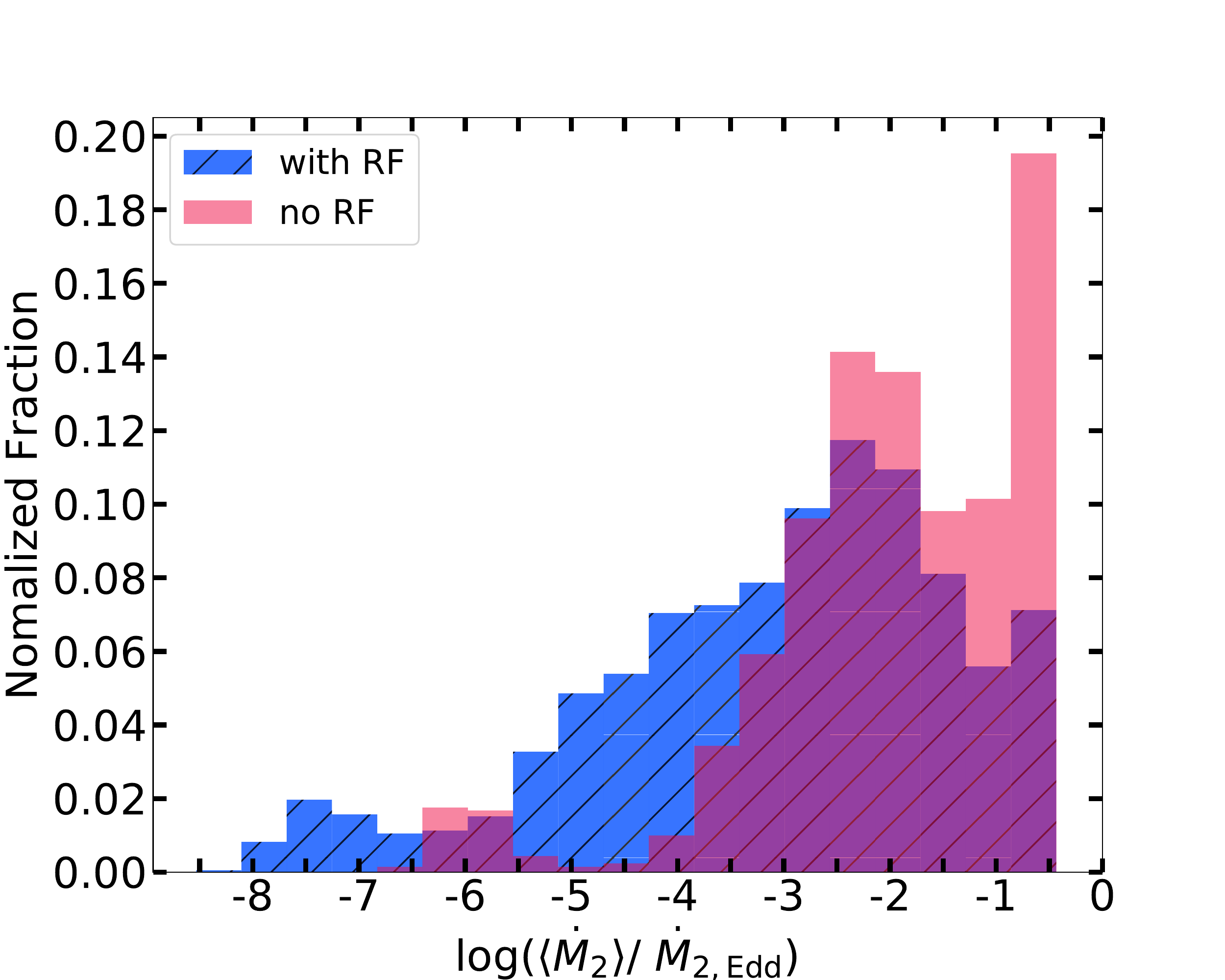}
\caption{The Eddington factor distribution of sMBHs with/ without radiation feedback effect.} 
\label{fig:m2_dot}
\end{figure*}

Without radiation feedback the distribution of the time-averaged accretion rates onto the sMBHs peaks sharply at an Eddington ratio of $10^{\rm -0.5}$ with a median of $10^{\rm -1.8}$ (Figure~\ref{fig:m2_dot}). The time-averaged accretion rate onto the pMBHs is tightly clustered around the upper limit of $10^{\rm -1}$ Eddington (equation~(\ref{eq:l1})) with a median Eddington ratio of $10^{\rm -1}$. After including the effects of radiation feedback, the sMBHs have a time-averaged accretion rate distribution that peaks at an Eddington ratio of $10^{\rm -2.5}$ with a median of  $10^{\rm -2.7}$ (Figure~\ref{fig:m2_dot}). The accretion rate distribution of pMBHs is almost unchanged with radiation feedback.

\begin{thebibliography}{}
\expandafter\ifx\csname natexlab\endcsname\relax\def\natexlab#1{#1}\fi
\providecommand{\url}[1]{\href{#1}{#1}}

\bibitem[{{Amaro-Seoane} {et~al.}(2017){Amaro-Seoane}, {Audley}, {Babak},
  {Baker}, {Barausse}, {Bender}, {Berti}, {Binetruy}, {Born}, {Bortoluzzi},
  {Camp}, {Caprini}, {Cardoso}, {Colpi}, {Conklin}, {Cornish}, {Cutler},
  {Danzmann}, {Dolesi}, {Ferraioli}, {Ferroni}, {Fitzsimons}, {Gair}, {Gesa
  Bote}, {Giardini}, {Gibert}, {Grimani}, {Halloin}, {Heinzel}, {Hertog},
  {Hewitson}, {Holley-Bockelmann}, {Hollington}, {Hueller}, {Inchauspe},
  {Jetzer}, {Karnesis}, {Killow}, {Klein}, {Klipstein}, {Korsakova}, {Larson},
  {Livas}, {Lloro}, {Man}, {Mance}, {Martino}, {Mateos}, {McKenzie},
  {McWilliams}, {Miller}, {Mueller}, {Nardini}, {Nelemans}, {Nofrarias},
  {Petiteau}, {Pivato}, {Plagnol}, {Porter}, {Reiche}, {Robertson},
  {Robertson}, {Rossi}, {Russano}, {Schutz}, {Sesana}, {Shoemaker}, {Slutsky},
  {Sopuerta}, {Sumner}, {Tamanini}, {Thorpe}, {Troebs}, {Vallisneri},
  {Vecchio}, {Vetrugno}, {Vitale}, {Volonteri}, {Wanner}, {Ward}, {Wass},
  {Weber}, {Ziemer}, \& {Zweifel}}]{LISA2017}
{Amaro-Seoane}, P., {Audley}, H., {Babak}, S., {et~al.} 2017, arXiv e-prints,
  arXiv:1702.00786

\bibitem[{{Antonini} \& {Merritt}(2012)}]{AM2012}
{Antonini}, F., \& {Merritt}, D. 2012, \apj, 745, 83

\bibitem[{{Barnes} \& {Hernquist}(1991)}]{Barnes1991}
{Barnes}, J.~E., \& {Hernquist}, L.~E. 1991, \apjl, 370, L65

\bibitem[{{Begelman}(1985)}]{B1985}
{Begelman}, M.~C. 1985, \apj, 297, 492

\bibitem[{{Begelman} {et~al.}(1980){Begelman}, {Blandford}, \&
  {Rees}}]{BBR1980}
{Begelman}, M.~C., {Blandford}, R.~D., \& {Rees}, M.~J. 1980, \nat, 287, 307

\bibitem[{{Binney} \& {Tremaine}(2008)}]{BT1987}
{Binney}, J., \& {Tremaine}, S. 2008, {Galactic Dynamics: Second Edition}
  (Princeton University Press)

\bibitem[{{Bondi}(1952)}]{Bondi1952}
{Bondi}, H. 1952, \mnras, 112, 195

\bibitem[{{Bondi} \& {Hoyle}(1944)}]{BH1944}
{Bondi}, H., \& {Hoyle}, F. 1944, \mnras, 104, 273

\bibitem[{{Burke-Spolaor}(2011)}]{BS2011}
{Burke-Spolaor}, S. 2011, \mnras, 410, 2113

\bibitem[{{Capelo} {et~al.}(2017{\natexlab{a}}){Capelo}, {Dotti}, {Volonteri},
  {Mayer}, {Bellovary}, \& {Shen}}]{Cap2017}
{Capelo}, P.~R., {Dotti}, M., {Volonteri}, M., {et~al.} 2017{\natexlab{a}},
  \mnras, 469, 4437

\bibitem[{{Capelo} {et~al.}(2017{\natexlab{b}}){Capelo}, {Dotti}, {Volonteri},
  {Mayer}, {Bellovary}, \& {Shen}}]{Capelo2017}
---. 2017{\natexlab{b}}, \mnras, 469, 4437

\bibitem[{{Comerford} {et~al.}(2012){Comerford}, {Gerke}, {Stern}, {Cooper},
  {Weiner}, {Newman}, {Madsen}, \& {Barrows}}]{C2012}
{Comerford}, J.~M., {Gerke}, B.~F., {Stern}, D., {et~al.} 2012, \apj, 753, 42

\bibitem[{{Crenshaw} {et~al.}(2010){Crenshaw}, {Schmitt}, {Kraemer},
  {Mushotzky}, \& {Dunn}}]{Crenshaw2010}
{Crenshaw}, D.~M., {Schmitt}, H.~R., {Kraemer}, S.~B., {Mushotzky}, R.~F., \&
  {Dunn}, J.~P. 2010, \apj, 708, 419

\bibitem[{{De Rosa} {et~al.}(2019){De Rosa}, {Vignali}, {Bogdanovi{\'c}},
  {Capelo}, {Charisi}, {Dotti}, {Husemann}, {Lusso}, {Mayer}, {Paragi},
  {Runnoe}, {Sesana}, {Steinborn}, {Bianchi}, {Colpi}, {del Valle}, {Frey},
  {Gab{\'a}nyi}, {Giustini}, {Guainazzi}, {Haiman}, {Herrera Ruiz},
  {Herrero-Illana}, {Iwasawa}, {Komossa}, {Lena}, {Loiseau}, {Perez-Torres},
  {Piconcelli}, \& {Volonteri}}]{Rosa2020}
{De Rosa}, A., {Vignali}, C., {Bogdanovi{\'c}}, T., {et~al.} 2019, \nar, 86,
  101525

\bibitem[{{Di Matteo} {et~al.}(2005){Di Matteo}, {Springel}, \&
  {Hernquist}}]{D2005}
{Di Matteo}, T., {Springel}, V., \& {Hernquist}, L. 2005, \nat, 433, 604

\bibitem[{{Foord} {et~al.}(2020){Foord}, {G{\"u}ltekin}, {Nevin}, {Comerford},
  {Hodges-Kluck}, {Barrows}, {Goulding}, \& {Greene}}]{Foord2020}
{Foord}, A., {G{\"u}ltekin}, K., {Nevin}, R., {et~al.} 2020, \apj, 892, 29

\bibitem[{{Foord} {et~al.}(2019){Foord}, {G{\"u}ltekin}, {Reynolds},
  {Hodges-Kluck}, {Cackett}, {Comerford}, {King}, {Miller}, \&
  {Runnoe}}]{Foord2019}
{Foord}, A., {G{\"u}ltekin}, K., {Reynolds}, M.~T., {et~al.} 2019, \apj, 877,
  17

\bibitem[{{Fu} {et~al.}(2015{\natexlab{a}}){Fu}, {Myers}, {Djorgovski}, {Yan},
  {Wrobel}, \& {Stockton}}]{Fu2015a}
{Fu}, H., {Myers}, A.~D., {Djorgovski}, S.~G., {et~al.} 2015{\natexlab{a}},
  \apj, 799, 72

\bibitem[{{Fu} {et~al.}(2015{\natexlab{b}}){Fu}, {Wrobel}, {Myers},
  {Djorgovski}, \& {Yan}}]{Fu2015b}
{Fu}, H., {Wrobel}, J.~M., {Myers}, A.~D., {Djorgovski}, S.~G., \& {Yan}, L.
  2015{\natexlab{b}}, \apjl, 815, L6

\bibitem[{{Fu} {et~al.}(2012){Fu}, {Yan}, {Myers}, {Stockton}, {Djorgovski},
  {Aldering}, \& {Rich}}]{Fu2012}
{Fu}, H., {Yan}, L., {Myers}, A.~D., {et~al.} 2012, \apj, 745, 67

\bibitem[{{Gardner} {et~al.}(2006){Gardner}, {Mather}, {Clampin}, {Doyon},
  {Greenhouse}, {Hammel}, {Hutchings}, {Jakobsen}, {Lilly}, {Long}, {Lunine},
  {McCaughrean}, {Mountain}, {Nella}, {Rieke}, {Rieke}, {Rix}, {Smith},
  {Sonneborn}, {Stiavelli}, {Stockman}, {Windhorst}, \& {Wright}}]{jwst}
{Gardner}, J.~P., {Mather}, J.~C., {Clampin}, M., {et~al.} 2006, \ssr, 123, 485

\bibitem[{{Gross} {et~al.}(2019){Gross}, {Fu}, {Myers}, {Wrobel}, \&
  {Djorgovski}}]{Gross2019}
{Gross}, A.~C., {Fu}, H., {Myers}, A.~D., {Wrobel}, J.~M., \& {Djorgovski},
  S.~G. 2019, \apj, 883, 50

\bibitem[{{Gruzinov} {et~al.}(2020){Gruzinov}, {Levin}, \& {Matzner}}]{G2020}
{Gruzinov}, A., {Levin}, Y., \& {Matzner}, C.~D. 2020, \mnras, 492, 2755

\bibitem[{{Hopkins} {et~al.}(2012){Hopkins}, {Hayward}, {Narayanan}, \&
  {Hernquist}}]{Hop2012}
{Hopkins}, P.~F., {Hayward}, C.~C., {Narayanan}, D., \& {Hernquist}, L. 2012,
  \mnras, 420, 320

\bibitem[{{Hopkins} {et~al.}(2005{\natexlab{a}}){Hopkins}, {Hernquist}, {Cox},
  {Di Matteo}, {Martini}, {Robertson}, \& {Springel}}]{Hop2005}
{Hopkins}, P.~F., {Hernquist}, L., {Cox}, T.~J., {et~al.} 2005{\natexlab{a}},
  \apj, 630, 705

\bibitem[{{Hopkins} {et~al.}(2005{\natexlab{b}}){Hopkins}, {Hernquist},
  {Martini}, {Cox}, {Robertson}, {Di Matteo}, \& {Springel}}]{Hop2005_m}
{Hopkins}, P.~F., {Hernquist}, L., {Martini}, P., {et~al.} 2005{\natexlab{b}},
  \apjl, 625, L71

\bibitem[{{Hou} {et~al.}(2020){Hou}, {Li}, \& {Liu}}]{Hou2020}
{Hou}, M., {Li}, Z., \& {Liu}, X. 2020, \apj, 900, 79

\bibitem[{{Hou} {et~al.}(2019){Hou}, {Liu}, {Guo}, {Li}, {Shen}, \&
  {Green}}]{Hou2019}
{Hou}, M., {Liu}, X., {Guo}, H., {et~al.} 2019, \apj, 882, 41

\bibitem[{{Hoyle} \& {Lyttleton}(1939)}]{HL1939}
{Hoyle}, F., \& {Lyttleton}, R.~A. 1939, Proceedings of the Cambridge
  Philosophical Society, 35, 405

\bibitem[{{Inayoshi} {et~al.}(2016){Inayoshi}, {Haiman}, \& {Ostriker}}]{I2016}
{Inayoshi}, K., {Haiman}, Z., \& {Ostriker}, J.~P. 2016, \mnras, 459, 3738

\bibitem[{{Kelley} {et~al.}(2019){Kelley}, {Charisi}, {Burke-Spolaor}, {Simon},
  {Blecha}, {Bogdanovic}, {Colpi}, {Comerford}, {D'Orazio}, {Dotti},
  {Eracleous}, {Graham}, {Greene}, {Haiman}, {Holley-Bockelmann}, {Kara},
  {Kelly}, {Komossa}, {Larson}, {Liu}, {Ma}, {Noble}, {Paschalidis}, {Rafikov},
  {Ravi}, {Runnoe}, {Sesana}, {Stern}, {Strauss}, {U}, {Volonteri}, \&
  {Nanograv Collaboration}}]{Kelley2019}
{Kelley}, L., {Charisi}, M., {Burke-Spolaor}, S., {et~al.} 2019, \baas, 51, 490

\bibitem[{{Kelley} {et~al.}(2017){Kelley}, {Blecha}, \& {Hernquist}}]{KBH2017}
{Kelley}, L.~Z., {Blecha}, L., \& {Hernquist}, L. 2017, \mnras, 464, 3131

\bibitem[{{Kim} \& {Kim}(2007)}]{KK2007}
{Kim}, H., \& {Kim}, W.-T. 2007, \apj, 665, 432

\bibitem[{{Kormendy} \& {Ho}(2013)}]{Kormendy2013}
{Kormendy}, J., \& {Ho}, L.~C. 2013, \araa, 51, 511

\bibitem[{{Koss} {et~al.}(2012){Koss}, {Mushotzky}, {Treister}, {Veilleux},
  {Vasudevan}, \& {Trippe}}]{Koss2012}
{Koss}, M., {Mushotzky}, R., {Treister}, E., {et~al.} 2012, \apjl, 746, L22

\bibitem[{{Li} {et~al.}(2020{\natexlab{a}}){Li}, {Bogdanovi{\'c}}, \&
  {Ballantyne}}]{LBB2020}
{Li}, K., {Bogdanovi{\'c}}, T., \& {Ballantyne}, D.~R. 2020{\natexlab{a}},
  \apj, 896, 113 (LBB20)

\bibitem[{{Li} {et~al.}(2020{\natexlab{b}}){Li}, {Bogdanovi{\'c}}, \&
  {Ballantyne}}]{LBB20b}
---. 2020{\natexlab{b}}, \apj, 905, 123 (LBB20b)

\bibitem[{{Lusso} {et~al.}(2012){Lusso}, {Comastri}, {Simmons}, {Mignoli},
  {Zamorani}, {Vignali}, {Brusa}, {Shankar}, {Lutz}, {Trump}, {Maiolino},
  {Gilli}, {Bolzonella}, {Puccetti}, {Salvato}, {Impey}, {Civano}, {Elvis},
  {Mainieri}, {Silverman}, {Koekemoer}, {Bongiorno}, {Merloni}, {Berta}, {Le
  Floc'h}, {Magnelli}, {Pozzi}, \& {Riguccini}}]{Lusso2012}
{Lusso}, E., {Comastri}, A., {Simmons}, B.~D., {et~al.} 2012, \mnras, 425, 623

\bibitem[{{Magorrian} {et~al.}(1998){Magorrian}, {Tremaine}, {Richstone},
  {Bender}, {Bower}, {Dressler}, {Faber}, {Gebhardt}, {Green}, {Grillmair},
  {Kormendy}, \& {Lauer}}]{M1998}
{Magorrian}, J., {Tremaine}, S., {Richstone}, D., {et~al.} 1998, \aj, 115, 2285

\bibitem[{{McKinnon} {et~al.}(2019){McKinnon}, {Beasley}, {Murphy}, {Selina},
  {Farnsworth}, \& {Walter}}]{ngvla}
{McKinnon}, M., {Beasley}, A., {Murphy}, E., {et~al.} 2019, in Bulletin of the
  American Astronomical Society, Vol.~51, 81

\bibitem[{{Merloni} {et~al.}(2012){Merloni}, {Predehl}, {Becker},
  {B{\"o}hringer}, {Boller}, {Brunner}, {Brusa}, {Dennerl}, {Freyberg},
  {Friedrich}, {Georgakakis}, {Haberl}, {Hasinger}, {Meidinger}, {Mohr},
  {Nandra}, {Rau}, {Reiprich}, {Robrade}, {Salvato}, {Santangelo}, {Sasaki},
  {Schwope}, {Wilms}, \& {German eROSITA Consortium}}]{erosita}
{Merloni}, A., {Predehl}, P., {Becker}, W., {et~al.} 2012, arXiv e-prints,
  arXiv:1209.3114

\bibitem[{{Mezcua} {et~al.}(2014){Mezcua}, {Lobanov}, {Mediavilla}, \&
  {Karouzos}}]{Mez2014}
{Mezcua}, M., {Lobanov}, A.~P., {Mediavilla}, E., \& {Karouzos}, M. 2014, \apj,
  784, 16

\bibitem[{{M{\"u}ller-S{\'a}nchez} {et~al.}(2015){M{\"u}ller-S{\'a}nchez},
  {Comerford}, {Nevin}, {Barrows}, {Cooper}, \& {Greene}}]{Muller2015}
{M{\"u}ller-S{\'a}nchez}, F., {Comerford}, J.~M., {Nevin}, R., {et~al.} 2015,
  \apj, 813, 103

\bibitem[{{Nandra} {et~al.}(2013){Nandra}, {Barret}, {Barcons}, {Fabian}, {den
  Herder}, {Piro}, {Watson}, {Adami}, {Aird}, {Afonso}, {Alexander},
  {Argiroffi}, {Amati}, {Arnaud}, {Atteia}, {Audard}, {Badenes}, {Ballet},
  {Ballo}, {Bamba}, {Bhardwaj}, {Stefano Battistelli}, {Becker}, {De Becker},
  {Behar}, {Bianchi}, {Biffi}, {B{\^\i}rzan}, {Bocchino}, {Bogdanov}, {Boirin},
  {Boller}, {Borgani}, {Borm}, {Bouch{\'e}}, {Bourdin}, {Bower}, {Braito},
  {Branchini}, {Branduardi-Raymont}, {Bregman}, {Brenneman}, {Brightman},
  {Br{\"u}ggen}, {Buchner}, {Bulbul}, {Brusa}, {Bursa}, {Caccianiga},
  {Cackett}, {Campana}, {Cappelluti}, {Cappi}, {Carrera}, {Ceballos},
  {Christensen}, {Chu}, {Churazov}, {Clerc}, {Corbel}, {Corral}, {Comastri},
  {Costantini}, {Croston}, {Dadina}, {D'Ai}, {Decourchelle}, {Della Ceca},
  {Dennerl}, {Dolag}, {Done}, {Dovciak}, {Drake}, {Eckert}, {Edge}, {Ettori},
  {Ezoe}, {Feigelson}, {Fender}, {Feruglio}, {Finoguenov}, {Fiore}, {Galeazzi},
  {Gallagher}, {Gandhi}, {Gaspari}, {Gastaldello}, {Georgakakis},
  {Georgantopoulos}, {Gilfanov}, {Gitti}, {Gladstone}, {Goosmann}, {Gosset},
  {Grosso}, {Guedel}, {Guerrero}, {Haberl}, {Hardcastle}, {Heinz}, {Alonso
  Herrero}, {Herv{\'e}}, {Holmstrom}, {Iwasawa}, {Jonker}, {Kaastra}, {Kara},
  {Karas}, {Kastner}, {King}, {Kosenko}, {Koutroumpa}, {Kraft}, {Kreykenbohm},
  {Lallement}, {Lanzuisi}, {Lee}, {Lemoine-Goumard}, {Lobban}, {Lodato},
  {Lovisari}, {Lotti}, {McCharthy}, {McNamara}, {Maggio}, {Maiolino}, {De
  Marco}, {de Martino}, {Mateos}, {Matt}, {Maughan}, {Mazzotta}, {Mendez},
  {Merloni}, {Micela}, {Miceli}, {Mignani}, {Miller}, {Miniutti}, {Molendi},
  {Montez}, {Moretti}, {Motch}, {Naz{\'e}}, {Nevalainen}, {Nicastro}, {Nulsen},
  {Ohashi}, {O'Brien}, {Osborne}, {Oskinova}, {Pacaud}, {Paerels}, {Page},
  {Papadakis}, {Pareschi}, {Petre}, {Petrucci}, {Piconcelli}, {Pillitteri},
  {Pinto}, {de Plaa}, {Pointecouteau}, {Ponman}, {Ponti}, {Porquet}, {Pounds},
  {Pratt}, {Predehl}, {Proga}, {Psaltis}, {Rafferty}, {Ramos-Ceja}, {Ranalli},
  {Rasia}, {Rau}, {Rauw}, {Rea}, {Read}, {Reeves}, {Reiprich}, {Renaud},
  {Reynolds}, {Risaliti}, {Rodriguez}, {Rodriguez Hidalgo}, {Roncarelli},
  {Rosario}, {Rossetti}, {Rozanska}, {Rovilos}, {Salvaterra}, {Salvato}, {Di
  Salvo}, {Sanders}, {Sanz-Forcada}, {Schawinski}, {Schaye}, {Schwope},
  {Sciortino}, {Severgnini}, {Shankar}, {Sijacki}, {Sim}, {Schmid}, {Smith},
  {Steiner}, {Stelzer}, {Stewart}, {Strohmayer}, {Str{\"u}der}, {Sun}, {Takei},
  {Tatischeff}, {Tiengo}, {Tombesi}, {Trinchieri}, {Tsuru}, {Ud-Doula},
  {Ursino}, {Valencic}, {Vanzella}, {Vaughan}, {Vignali}, {Vink}, {Vito},
  {Volonteri}, {Wang}, {Webb}, {Willingale}, {Wilms}, {Wise}, {Worrall},
  {Young}, {Zampieri}, {In't Zand}, {Zane}, {Zezas}, {Zhang}, \&
  {Zhuravleva}}]{athena}
{Nandra}, K., {Barret}, D., {Barcons}, X., {et~al.} 2013, arXiv e-prints,
  arXiv:1306.2307

\bibitem[{{Narayanan} {et~al.}(2008){Narayanan}, {Cox}, {Robertson},
  {Dav{\'e}}, {Di Matteo}, {Hernquist}, {Hopkins}, {Kulesa}, \&
  {Walker}}]{Nara2008}
{Narayanan}, D., {Cox}, T.~J., {Robertson}, B., {et~al.} 2008, in Astronomical
  Society of the Pacific Conference Series, Vol. 381, Infrared Diagnostics of
  Galaxy Evolution, ed. R.~R. {Chary}, H.~I. {Teplitz}, \& K.~{Sheth}, 491

\bibitem[{{Ostriker}(1999)}]{O1999}
{Ostriker}, E.~C. 1999, \apj, 513, 252

\bibitem[{{Ostriker} {et~al.}(1976){Ostriker}, {McCray}, {Weaver}, \&
  {Yahil}}]{Ostriker1976}
{Ostriker}, J.~P., {McCray}, R., {Weaver}, R., \& {Yahil}, A. 1976, \apjl, 208,
  L61

\bibitem[{{Park} \& {Bogdanovi{\'c}}(2017)}]{PB2017}
{Park}, K., \& {Bogdanovi{\'c}}, T. 2017, \apj, 838, 103

\bibitem[{{Park} \& {Ricotti}(2011)}]{Park2011}
{Park}, K., \& {Ricotti}, M. 2011, \apj, 739, 2

\bibitem[{{Park} \& {Ricotti}(2012)}]{Park2012}
---. 2012, \apj, 747, 9

\bibitem[{{Park} \& {Ricotti}(2013)}]{Park2013}
---. 2013, \apj, 767, 163

\bibitem[{{Prandoni} \& {Seymour}(2015)}]{ska}
{Prandoni}, I., \& {Seymour}, N. 2015, in Advancing Astrophysics with the
  Square Kilometre Array (AASKA14), 67

\bibitem[{{Ricotti} {et~al.}(2008){Ricotti}, {Ostriker}, \& {Mack}}]{Ric2008}
{Ricotti}, M., {Ostriker}, J.~P., \& {Mack}, K.~J. 2008, \apj, 680, 829

\bibitem[{{Rosario} {et~al.}(2010){Rosario}, {Shields}, {Taylor}, {Salviand
  er}, \& {Smith}}]{Rosario2010}
{Rosario}, D.~J., {Shields}, G.~A., {Taylor}, G.~B., {Salviand er}, S., \&
  {Smith}, K.~L. 2010, \apj, 716, 131

\bibitem[{{Rosas-Guevara} {et~al.}(2019){Rosas-Guevara}, {Bower}, {McAlpine},
  {Bonoli}, \& {Tissera}}]{Rosas2019}
{Rosas-Guevara}, Y.~M., {Bower}, R.~G., {McAlpine}, S., {Bonoli}, S., \&
  {Tissera}, P.~B. 2019, \mnras, 483, 2712

\bibitem[{{Sargent} {et~al.}(1989){Sargent}, {Sanders}, \&
  {Phillips}}]{Sar1989}
{Sargent}, A.~I., {Sanders}, D.~B., \& {Phillips}, T.~G. 1989, \apjl, 346, L9

\bibitem[{{Sargent} {et~al.}(1987){Sargent}, {Sanders}, {Scoville}, \&
  {Soifer}}]{Sar1987}
{Sargent}, A.~I., {Sanders}, D.~B., {Scoville}, N.~Z., \& {Soifer}, B.~T. 1987,
  \apjl, 312, L35

\bibitem[{{Scoville} {et~al.}(1986){Scoville}, {Sanders}, {Sargent}, {Soifer},
  {Scott}, \& {Lo}}]{Sco1986}
{Scoville}, N.~Z., {Sanders}, D.~B., {Sargent}, A.~I., {et~al.} 1986, \apjl,
  311, L47

\bibitem[{{Severgnini} {et~al.}(2020){Severgnini}, {Braito}, {Cicone},
  {Saracco}, {Vignali}, {Serafinelli}, {Della Ceca}, {Dotti}, {Cusano},
  {Paris}, {Pruto}, {Zaino}, {Ballo}, \& {Landoni}}]{Severgnini2021}
{Severgnini}, P., {Braito}, V., {Cicone}, C., {et~al.} 2020, arXiv e-prints,
  arXiv:2012.09184

\bibitem[{{Springel} {et~al.}(2005){Springel}, {Di Matteo}, \&
  {Hernquist}}]{Spring2005}
{Springel}, V., {Di Matteo}, T., \& {Hernquist}, L. 2005, \mnras, 361, 776

\bibitem[{{Teng} {et~al.}(2012){Teng}, {Schawinski}, {Urry}, {Darg}, {Kaviraj},
  {Oh}, {Bonning}, {Cardamone}, {Keel}, {Lintott}, {Simmons}, \&
  {Treister}}]{Teng2012}
{Teng}, S.~H., {Schawinski}, K., {Urry}, C.~M., {et~al.} 2012, \apj, 753, 165

\bibitem[{{Toomre}(1964)}]{T1964}
{Toomre}, A. 1964, \apj, 139, 1217

\bibitem[{{Toyouchi} {et~al.}(2020){Toyouchi}, {Hosokawa}, {Sugimura}, \&
  {Kuiper}}]{T2020}
{Toyouchi}, D., {Hosokawa}, T., {Sugimura}, K., \& {Kuiper}, R. 2020, \mnras,
  arXiv:2002.08017

\bibitem[{{Van Wassenhove} {et~al.}(2012){Van Wassenhove}, {Volonteri},
  {Mayer}, {Dotti}, {Bellovary}, \& {Callegari}}]{VanW2012}
{Van Wassenhove}, S., {Volonteri}, M., {Mayer}, L., {et~al.} 2012, \apjl, 748,
  L7

\bibitem[{{White} \& {Frenk}(1991)}]{White1991}
{White}, S. D.~M., \& {Frenk}, C.~S. 1991, \apj, 379, 52

\bibitem[{{White} \& {Rees}(1978)}]{White1978}
{White}, S.~D.~M., \& {Rees}, M.~J. 1978, \mnras, 183, 341

\end{thebibliography}
\end{document}